\documentclass[aps,pra,twocolumn,amsmath,amssymb,nofootinbib,superscriptaddress,floatfix,reprint,longbibliography]{revtex4-2}
\usepackage[dvips]{graphicx}
\usepackage{latexsym}
\usepackage{amsmath}
\usepackage{amsfonts}
\usepackage{amssymb}
\usepackage{bm}
\usepackage{color}
\usepackage{txfonts}
\usepackage{float}
\usepackage{url}\href{}{}
\usepackage[version=4]{mhchem}
\usepackage{braket}
\usepackage{booktabs}
\usepackage{siunitx}
\usepackage{mhchem}
\usepackage[colorlinks=true, urlcolor=blue, linkcolor=blue, citecolor=blue, pdftex]{hyperref}
\usepackage{ulem}
\makeatletter

\newcommand{\Rmnum}[1]{\expandafter\@slowromancap\romannumeral #1@}
\makeatother

\begin{document}

\title{Optimizing attosecond pulse generation in solids by modulating electronic dynamics with monochromatic laser field}

\author{Xinyuan Zhang}
\thanks{These authors contributed equally to this work.}
\affiliation{Beijing National Laboratory for Condensed Matter Physics and Institute of Physics, Chinese Academy of Sciences, Beijing 100190, China}
\affiliation{School of Physical Sciences, University of Chinese Academy of Sciences, Beijing 100190, China}

\author{Shiqi Hu}
\thanks{These authors contributed equally to this work.}
\affiliation{Beijing National Laboratory for Condensed Matter Physics and Institute of Physics, Chinese Academy of Sciences, Beijing 100190, China}
\affiliation{School of Physical Sciences, University of Chinese Academy of Sciences, Beijing 100190, China}

\author{Mengxue Guan}
\email{mxguan@bit.edu.cn}
\affiliation{Centre for Quantum Physics, Key Laboratory of Advanced Optoelectronic Quantum Architecture and Measurement (Ministry of Education), School of Physics, Beijing Institute of Technology, Beijing 100081, China}
\affiliation{Beijing National Laboratory for Condensed Matter Physics and Institute of Physics, Chinese Academy of Sciences, Beijing 100190, China}

\author{Sheng Meng}
\email{smeng@iphy.ac.cn}
\affiliation{Beijing National Laboratory for Condensed Matter Physics and Institute of Physics, Chinese Academy of Sciences, Beijing 100190, China}
\affiliation{School of Physical Sciences, University of Chinese Academy of Sciences, Beijing 100190, China}
\affiliation{Songshan Lake Materials Laboratory, Dongguan, Guangdong 523808, China}

\begin{abstract}
A practical approach is proposed for efficiently generating ultrashort attosecond pulses (APs) from realistic solid-state materials, aiming to optimize pulse width effectively. By adjusting the photon energy while maintaining a constant peak electric field, this strategy modulates the peak vector potential and laser field period, thereby controlling the high harmonic cutoff energy and the time-domain emission characteristics of the harmonics. The field-driven electronic dynamics lead to a non-monotonic variation in both the intensity and duration of the generated attosecond pulses. The light field frequency can be adjusted to yield the optimal pulse. Beyond the primary demonstration with hexagonal boron nitride as a prototypical material, significant pulse width optimization has been achieved across a range of different materials. This straightforward and versatile strategy shows promise for application in solid-state materials, offering new pathways to promote high harmonic performance.
\end{abstract}
\maketitle

\section{Introduction}
Recent advancements in attosecond metrology \cite{mashiko2016petahertz,schotz2019perspective,fohlisch2005direct,schultze2010delay,cavalieri2007attosecond,sansone2012electron} have enabled the exploration of matter dynamics on increasingly shorter timescales. The duration of attosecond pulses (APs) limits temporal resolution, with the shortest AP of 43 attoseconds achieved in inert gases via high harmonic generation (HHG) \cite{gaumnitz2017streaking}. While HHG in gases is well understood through the semi-classical three-step model \cite{corkum1993plasma,schafer1993above,lewenstein1994theory}, HHG from solid-state materials, demonstrated since 2011, presents unique challenges and opportunities in ultrafast science \cite{ghimire2011observation,ghimire2012generation,ndabashimiye2016solid,liu2017high,you2017laser,wu2016multilevel,wu2015high,guan2016high,ikemachi2017trajectory,tancogne2017impact,tancogne2017ellipticity,tancogne2018atomic,guan2019cooperative,du2017quasi,ghimire2019high}. The harmonic emission in solids relies on their atomic and electronic structures, allowing HHG to serve as an all-optical method for probing intrinsic properties of materials \cite{vampa2015all,hu2024solid,luu2018measurement,lakhotia2020laser,hu2024phonon}. However, generating extremely short APs comparable to those from gases remains challenging \cite{li2020attosecond}, mainly due to complex electron dynamics, including interband electronic excitation and intraband electronic motion \cite{du2017quasi,ghimire2019high}.

APs are typically extracted from the plateau region of HHG spectra using amplitude gating methods \cite{corkum2007attosecond,chini2014generation,krausz2014attosecond,calegari2016advances,li2020attosecond}, highlighting the significance of plateau morphology \cite{li2017enhancement}. Solid-state materials, with their band structures and higher electron densities, exhibit multiple plateaus and larger cutoff energies ($\varepsilon_c$) in the HHG spectrum compared to gases \cite{ndabashimiye2016solid, tancogne2017impact}. The highest plateau is attributed to electron dynamics in high-lying bands \cite{wu2015high}, modifiable by adjusting the laser waveform. The vector potential $\mathbf{A}(t)$ of the laser field drives the ponderomotive motion of electrons \cite{kruchinin2018colloquium}, determining their range of motion within the energy bands and the recombination energy, which tunes the HHG spectrum's cutoff energy \cite{wu2016multilevel, li2017enhancement}. However, accurately modeling carrier dynamics in real solids remains challenging due to complex band structures, particularly in high-energy bands.

Here we aim to systematically design laser waveforms to modulate HHG and optimize AP generation from the highest plateau region in solid-state materials. Our investigation employs first-principles time-dependent density functional theory (TDDFT) simulations \cite{runge1984density,van1998causality}. Using monolayer hexagonal boron nitride (1L-hBN) with wide band gap as a prototype, we demonstrate that a heightened peak vector potential ($A_{p}$) promotes electronic excitation and intraband motion, resulting in broader electron occupation of high-energy conduction bands and extending the plateau region with a larger cutoff energy. Meanwhile, the oscillation period (\textit{T}) also modulates harmonic emission due to attosecond chirp effects \cite{wu2015high,you2017laser}. The influence of the external field on the harmonic plateau region and cutoff energy, along with the field induced harmonic chirp, collectively determines the width and intensity of ultrashort pulses. Based on this mechanism, we can achieve the generation of a 143 attosecond pulse theoretically, surpassing existing theoretical results for solid-state materials \cite{wu2021enhancement,guan2020toward,nourbakhsh2021high,sadeghifaraz2022efficient}.

\section{SIMULATION SETUP AND FIELD CONFIGURATION}
We focus on HHG properties with laser polarization along the armchair direction (M$_2$-$\Gamma$-M$_1$ line in reciprocal space) (Fig. \ref{fig;structure HHG heatmap}(a)) using TDDFT simulations \cite{ullrich2011time,tancogne2020octopus}. The field application time is 53.4 fs, and the waveform is given by
\begin{align}
E\left( t \right)=E_{p}g\left( t \right)\cos\left( \omega \left ( t-t_{0}  \right ) \right),
\end{align} 
where $E_{p}$ represents the peak electric field strength of 0.53 V/$\AA$, fixed at $t_0=26.7$ fs, and $g\left( t \right)=e^{-\frac{\left( t-t_{0} \right)^{2}}{2\sigma^{2} }}$ corresponds to a Gaussian-shaped envelope centered at $t_0$ with $\sigma=4.5$ fs. The photon energy $\hbar\omega$ is adjusted to enable continuous tuning of the vector potential  $\mathbf{A}\left( t \right)=-\int \mathbf{E}\left( t \right)\mathrm{d}t$, with peak value $A_{p}$ increasing monotonically as $\hbar\omega$ decreases (Fig. \ref{fig;structure HHG heatmap}(b)). It should be noted that the difference between the initial and final values of the vector potential, obtained through integration, is small enough to not affect the results. The mirror symmetry of 1L-hBN limits the current emission occurring only along the incident laser polarization direction \cite{liu2017high}. To mitigate the dipole moment contribution from residual electron occupation in conduction bands (CBs) and valance bands (VBs) after the optical field ends, the simulated current density is multiplied by the pulse envelope $g(t)$ to improve the signal-to-noise ratio \cite{wu2015high,guan2016high,li2017enhancement,ikemachi2017trajectory}. Further details can be found in Appendix \ref{app:TDDFT}.

\section{results and discussion}
\subsection{HHG spectra under different laser frequencies}
\begin{figure}[!htbp]
\centering
\includegraphics[width=8.6cm]{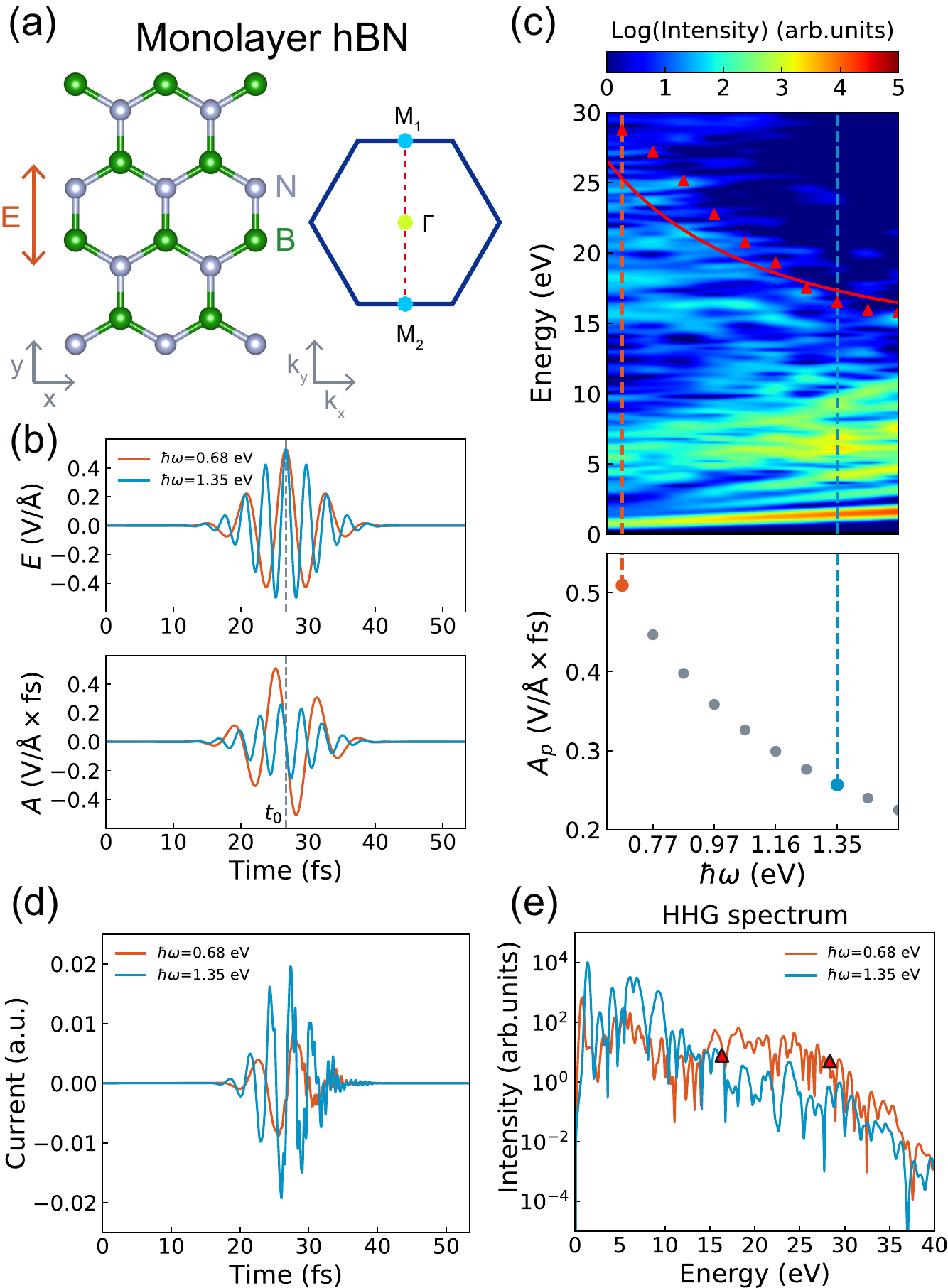}
\caption{(a) Crystal structure and the first Brillouin zone of 1L-hBN. The electric field of the laser pulse is polarized along the armchair direction. (b) The electric field and vector potential for photon energies of 0.68 eV and 1.35 eV, respectively. At $t_0=26.7$ fs, the electric field reaches its peak value, and the vector potential is zero. (c) The harmonic intensity and $A_{p}$ as a function of photon energy. Vertical dashed lines highlight the two cases displayed in (b). The triangle markers in the heatmap indicate $\varepsilon_c$ of TDDFT results, while the red curve representing $\varepsilon_c$ calculated by the four-level model. (d) and (e) Laser induced currents and HHG spectra for the two representative photon energies, where the triangle markers indicate the corresponding $\varepsilon_c$.}\label{fig;structure HHG heatmap}
\end{figure}

Figures \ref{fig;structure HHG heatmap}(c-e) depict the calculated current along the armchair direction and the corresponding harmonics emission spectra. Due to the narrower gaps compared to those in atomic gases, the HHG spectra do not exhibit a sharp decline as the harmonic energy increases, posing challenges in accurately determining $\varepsilon_c$ \cite{guan2020toward}. However, there is a clear trend where the approximate $\varepsilon_c$ increases as $\hbar\omega$ decreases red dots in Fig. \ref{fig;structure HHG heatmap}(c)), consistent with previous findings showing a positive correlation between $\varepsilon_c$ and $A_{p}$ \cite{li2017enhancement,wu2016multilevel}.

\begin{figure}[!htbp]
\centering
\includegraphics[width=8.6cm]{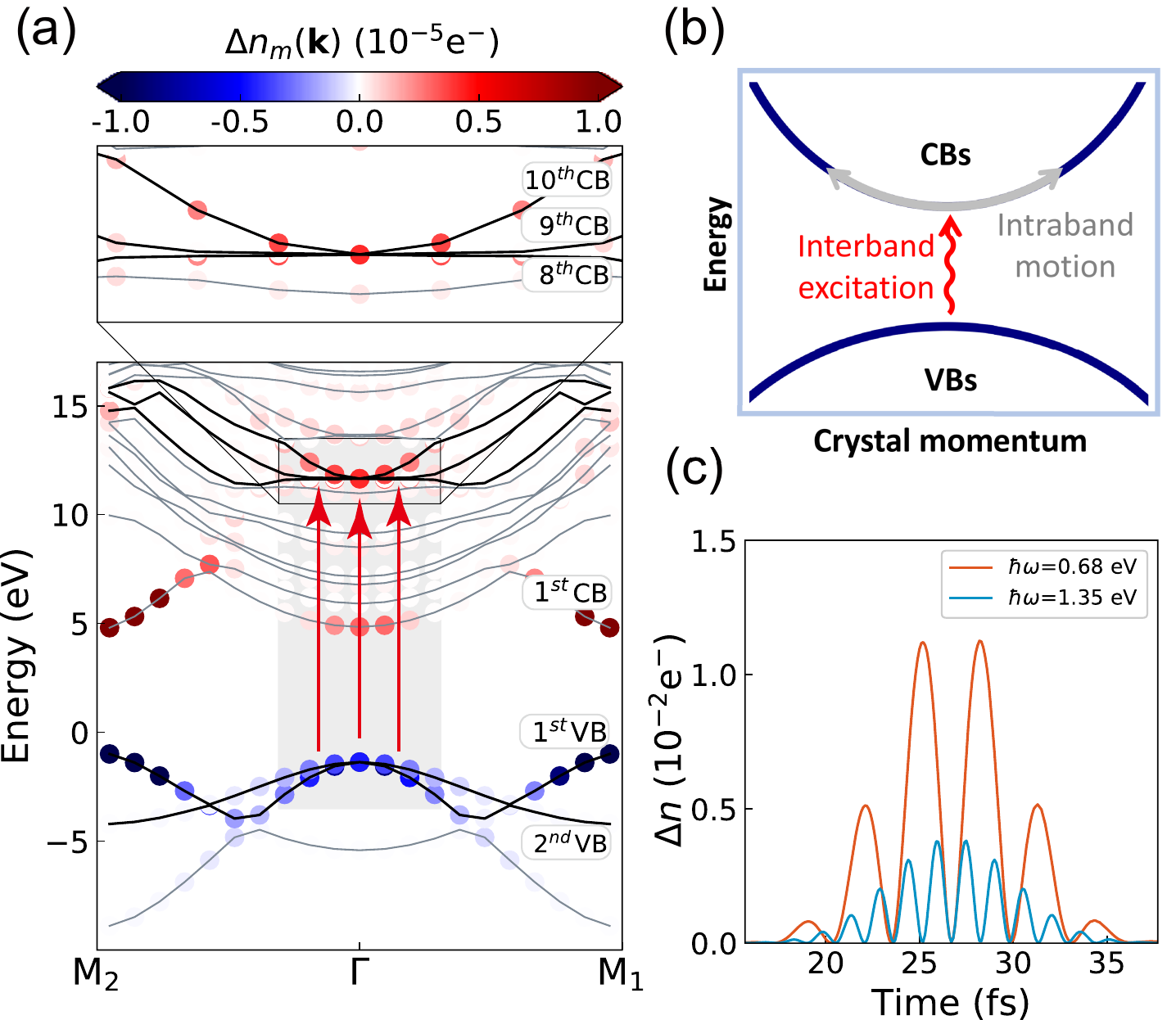}
\caption{(a) Changes in electron occupancy $\Delta n_{m}\left( \mathbf{k} \right)$ along M$_2$-$\Gamma$-M$_1$ line when $A( t_0)=0$. The red arrows indicate electronic transitions leading to the formation of the high-energy region in the harmonic spectrum. The inset amplifies the band structure around the $\Gamma$ point, focusing on the dominant CBs. (b) Schematic of interband excitation and intraband motion in reciprocal space. (c) Total number of excited electrons in the 8$^{th}$-10$^{th}$ conduction bands when photon energy is 0.68 eV and 1.35 eV, respectively.}\label{fig;direct excitation} 
\end{figure}

Figure \ref{fig;direct excitation}(a) shows calculated carrier occupation at $t_0$, focusing solely on the interband electronic excitation effect. At this given moment, $\textbf{A}\left( t_0 \right)=0$, so there is absence of the reciprocal space motion of the electron wavepacket, based on the acceleration theorem $\hbar \Delta \textbf{k}\left( t \right)=e\textbf{A}\left( t \right)$ \cite{rossi2002theory}. In the energy range near $\varepsilon_c$, the primary carrier excitation occurs around the $\Gamma$ point, involving transitions from the 1$^{st}$-2$^{nd}$ VBs to the 8$^{th}$-9$^{th}$ CBs (note CBs are labeled according to their energy in the ascending order starting from the Fermi level, while VBs are labeled in the descending energy order). Consequently, a model hamiltonian  within the subspace of these four Bloch wavefunctions at the $\Gamma$ point can be constructed to qualitatively illustrate the dependence of $\varepsilon_c$ on $A_p$, or $\hbar\omega$ in our current work. (details can be seen in Appendix \ref{app:four}). As illustrated in Fig. \ref{fig;structure HHG heatmap}(c), the four-level model and TDDFT simulations give consistent results.

\subsection{Interband and intraband electronic dynamics}
Let's first discuss how the vector potential field affects interband and intraband dynamics. The interband electronic dynamics during harmonic generation are explored by computing the time-dependent electron occupancy of CBs using TDDFT:
\begin{align}
\Delta n_{m}\left( \mathbf{k},t \right)=\frac{1}{N}\sum_{l}^{occ} \left| \left \langle \phi_{l,\mathbf{k}}\left(  \mathbf{r},t \right)  | \phi_{m,\mathbf{k}}\left( \mathbf{r},0 \right)  \right \rangle \right|^{2},
\end{align}
where $N$ is the number of \textbf{\textit{k}}-points, and $\phi_{m,\mathbf{k}}\left( \mathbf{r} ,t \right)$ represents the time-dependent Kohn-Sham wavefunction at the \textbf{\textit{k}}-point of the $m^{th}$ band in the first Brillouin zone. 
By summing $\Delta n_{m}\left( \mathbf{k},t \right)$ over significant excited bands, we obtain the electronic distribution $\Delta n\left( \mathbf{k},t \right)=\sum_m\Delta n_{m}\left( \mathbf{k},t \right)$ at a given \textbf{\textit{k}}. Further summing $\Delta n\left( \mathbf{k},t \right)$ over momentum space gives the total number of excited electrons $\Delta n\left( t \right)$. Figure \ref{fig;direct excitation}(c) illustrates the evolution of $\Delta n\left( t \right)$ occupying the 8$^{th}$-10$^{th}$ CBs around $\Gamma$. When the photon energy is 0.68 eV, $\Delta n$ exhibits a higher peak compared to the case of 1.35 eV, indicating that a larger $A{_p}$ field enhances interband excitation, thereby increasing the cutoff energy and harmonic intensity of the plateau region.

Based on time-dependent perturbation theory, we can understand the dependence of interband electronic excitation on $A_{p}$. The electronic transition probabilities from band $n_0$ to band $n$ at \textit{\textbf{k}}-point are expressed as: 
\begin{align}
    P_{n_{0}\to n}^{\mathbf{k}}\left (t\right ) = \left( E_{n}^{\mathbf{k}}-E_{n_{0}}^{\mathbf{k}} \right)^2 \left|\left\langle u_{n}^{\mathbf{k}} \right| \nabla_\mathbf{\mathbf{k}} \left| u_{n_{0}}^{\mathbf{k}} \right\rangle_{cell} \cdot \int_{0}^{t} \mathbf{A}\left ( t \right )  e^{i\left( E_{n}^{\mathbf{k}}-E_{n_{0}}^{\mathbf{k}} \right) t} dt \right|^{2}, \label{transition_probability}
\end{align}
the influence of the vector potential on the transition probability can be seen from the time-dependent part in Eq. \ref{transition_probability}. Given that the photon energy is much smaller compared to the transition energy differences ($ E_{n}^{\mathbf{k}}-E_{n_{0}}^{\mathbf{k}}$), $\textbf{A}\left(t\right)$ determines the envelope shape and amplitude of the function after integration. Therefore, an increase in $A_{p}$ results in a larger transition probability amplitude, indicating higher electron occupation. A more detailed analysis can be found in the Appendix \ref{app:tdpt}. Moreover, by calculating the dependence of overall transition probability on the transition energy difference $\Delta E$, an accurate energy threshold for dominant electronic transitions can be established. The corresponding energy range $\Delta E\geq 12.9$ eV aligns with transitions around the high-energy plateau of the HHG spectra (as indicated by the red arrows in Fig. \ref{fig;direct excitation}(a)).

\begin{figure*}[!htbp]
\centering
\includegraphics[width=0.9\textwidth]{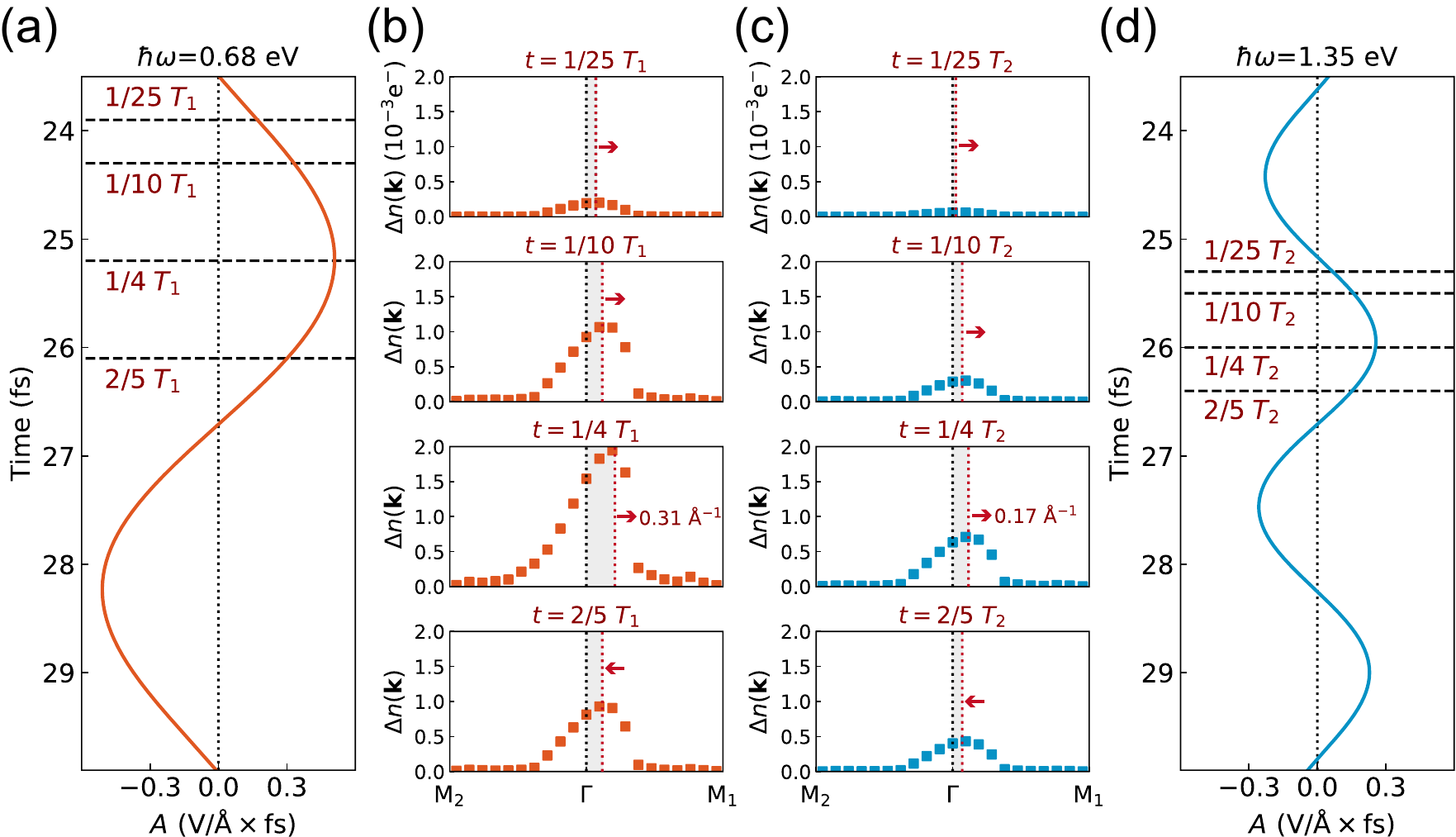}
\caption{(a) and (d) are the vector potentials at the photon energy of 0.68 eV and 1.35 eV, respectively. (b) and (c) show the electronic distribution on the 8$^{th}$-10$^{th}$ conduction bands along the M$_2$-$\Gamma$-M$_1$ line at four representative moments described in (a) and (d). Square markers illustrate changes in electron occupancy $\Delta n \left( \mathbf{k} \right)$. The gray shadows indicate the  range of electron motion.}\label{fig;wave packets oscillations}
\end{figure*}

In addition to interband electronic excitation, intraband electronic motion, which involve scattering among different \textit{\textbf{k}}-points, are also influenced by varying the vector potential (Fig. \ref{fig;direct excitation}(b)). Since electrons occupying the 8$^{th}$-10$^{th}$ CBs play a dominant role in determining the high-energy region of the harmonic spectrum, the time-dependent distribution $\Delta n\left( \mathbf{k},t \right)$ of these electrons is analyzed along the M$_2$-$\Gamma$-M$_1$ high-symmetry line. Figures \ref{fig;wave packets oscillations}(b) and (c) display $\Delta n \left( \mathbf{k},t \right)$ at four representative moments corresponding to 1/25, 1/10, 1/4 and 2/5 of the field period $T$. The oscillations of electron wavepackets are characterized by the movement of the peak position of $\Delta n \left( \mathbf{k},t \right)$ in the reciprocal space. In both laser fields, as the magnitude of $\mathbf{A}(t)$ increases (decreases), the peak shift from $\Gamma$ towards M$_1$ (M$_2$) point. Particularly, when the vector potential reaches its maximum value at $T/4$, electrons exhibit the widest distribution in momentum space. Moreover, a larger $A_{p}$ (e.g., $\hbar\omega$=0.68 eV) enhances intraband motion by driving the carriers across a broader region, i.e., $\Delta \textbf{k} \propto A_{p}$.

\subsection{Attosecond-pulse generation and optimization}
The idea behind the proposed waveform-design strategy can be summarized based on the above analysis. In solids, the vector potential contributes to the interband excitation of electrons. Besides, the vector potential also drives the ponderomotive motion, or intraband motion in momentum space. An increase in $A_{p}$ promotes cooperative action in both processes for harmonic generation. Consequently, more electrons occupy Bloch states with a high energy and large momentum, leading to stronger photon emission across extended energy ranges that reinforce the plateau region near $\varepsilon_c$, facilitating the generation of attosecond pulses using this plateau region.

\begin{figure}[!htbp]
\centering
\includegraphics[width=8.6cm]{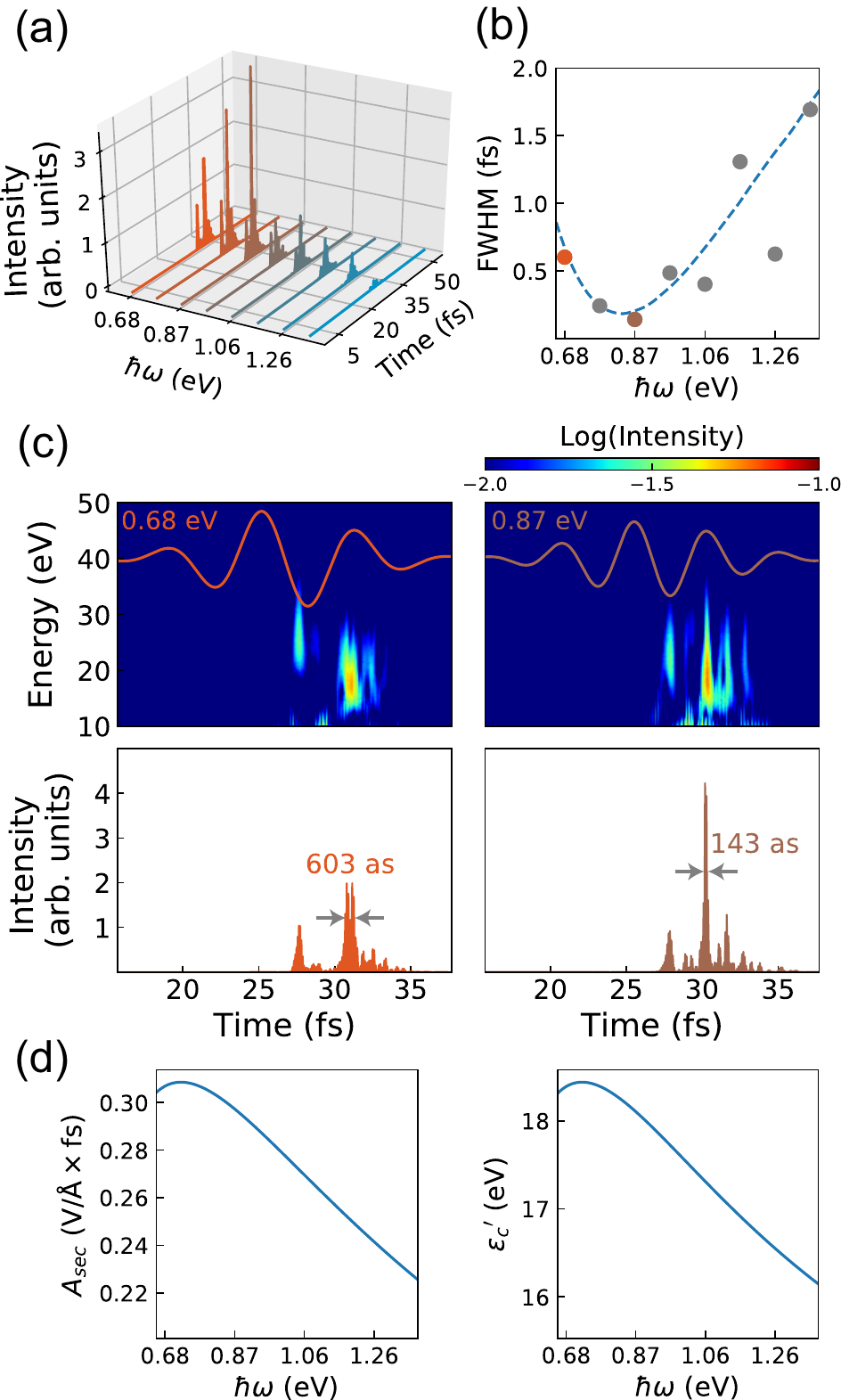}
\caption{(a) The ultrashort pulses obtained by filtering high harmonic spectra within the energy range of 15 eV to 60 eV at various photon energies. (b) The FWHM of the pulse as a function of photon energy. The blue dashed line represents a combined effect of laser frequency on the cutoff energy and the temporal emission width of the harmonic signal. (c) The upper graphs display the wavelet transform spectra for photon energies of 0.68 eV and 0.87 eV, respectively, overlaid with their corresponding vector potentials $\mathbf{A}(t)$. The lower graphs depict the generated ultrashort pulses. (d) Variation of the secondary peak of the vector potential $A_{sec}$ and the corresponding half-cycle cutoff energy ${\varepsilon_c}'$ with laser frequency.} \label{fig;pulse generate}
\end{figure} 

Figure \ref{fig;pulse generate}(a) displays ultrashort pulses obtained using amplitude gating methods, where high-energy plateau and cutoff regions (15 eV to 60 eV) of the harmonic spectra are filtered. However, there is a  discrepancy between the expected and actually computed pulse width. As $\hbar\omega$ decreases, the full width at half maximum (FWHM) of the generated pulse initially decreases and then increases, exhibiting non-monotonic behavior. This trend is highlighted by the dashed line serving as a guide in Fig. \ref{fig;pulse generate}(b). The optimal choice for generating the strongest and shortest pulse in 1L-hBN is $\hbar\omega= 0.87$ eV, achieving a FWHM of 143 as.

The non-monotonic variation trend of pulse width can be roughly understood through the previously mentioned mechanism, where the vector potential field regulates interband excitation and intraband motion of electrons, thereby controlling the harmonic plateau region and cutoff energy. Although the wavelet transform spectra at $\hbar\omega$ of 0.68 eV and 0.87 eV (Fig. \ref{fig;pulse generate}(c)) reveal that the main pulse appears at the positions of secondary peak of the vector potential, $A_{sec}$, under our field modulation, the secondary peak $A_{sec}$ still exhibits an approximately monotonic variation, with non-monotonic behavior emerging only at low frequencies as shown in Fig. \ref{fig;pulse generate}(d). By inputting $A_{sec}$ into the four-level model to obtain the half-cycle cutoff energy ${\varepsilon_c}'$, which changes monotonically in the medium and high frequencies, and exhibits non-monotonic behavior at lower frequencies. As a result, the pulse width generally increases with decreasing frequency but exhibits non-monotonicity at lower frequencies, basically consistent with Fig. \ref{fig;pulse generate}(b).

Additionally, comparing Figs. \ref{fig;pulse generate}(b) and (d) highlights a discrepancy in the locations of their extrema. This divergence arises from not considering the temporal emission characteristics of harmonics. Time-frequency analysis indicates that at $\hbar\omega= 0.68$ eV, the high-energy harmonics ($\geq 15$ eV) exhibit a broader emission width compared to those at $\hbar\omega= 0.87$ eV. This broadening is attributed to electron dynamics within adiabatic states, which create a time-dependent energy gap between photoexcited electrons and holes. Then, electron-hole recombination produces chirped signals with time-varying frequency components, mirroring the profile of the vector potential \cite{wu2015high,wu2016multilevel}. Increasing the optical field period $T$ will broadens the harmonic emission time range, as observed in the time-frequency spectra. As laser frequency decreases, this broadening becomes more pronounced, causing the pulse width extrema to shift rightward relative to those of the cutoff energy. Due to the presence of this effect, non-monotonic behavior is inevitably observed during pulse optimization. Therefore, by adjusting the optical field frequency, we can achieve the optimal pulse.

Although it is not theoretically possible to reduce the pulse width indefinitely, our approach still has its advantages. In recent years, innovative strategies have emerged to manipulate laser waveforms for attosecond pulse generation in theoretical works, including plasmonic fields and two-color fields. However, these strategies often require complex equipment and precise control \cite{wu2021enhancement,guan2020toward,nourbakhsh2021high,sadeghifaraz2022efficient}. The optimal attosecond pulse reported in our work is, on the one hand, narrower than those in previous theoretical studies, showcasing the potential of our pulse modulation approach. On the other hand, unlike those complex modulations, our approach relies on monochromatic light and achieves notable results with a laser peak intensity around $10^{12}$ W/cm$^2$, minimizing material damage \cite{roberts2011response,tancogne2018atomic,kong2022manipulation}. Furthermore, our modulation approach exhibits broad applicability, as it has been extended to modulate HHG in monolayer gallium selenide (GaSe) and Wurtzite-structured zinc oxide (ZnO) (see Appendix \ref{app:other}), the former being a two-dimensional material with a relatively large band gap \cite{rybkovskiy2011size}. The FWHM of the generated ultrashort pulses shows a similar non-monotonic dependence on photon energy, further demonstrating the versatility of the present mechanism across different materials.

However, due to the limitations of theoretical simulations, our approach uses an idealized numerical filter for pulse generation and does not account for the specific effects of experimental filters. Future studies aiming to incorporate the transmission function of experimental filters, such as an aluminum filter for our chosen energy window, would be a worthwhile direction and challenge, enabling more accurate estimations of pulse durations under realistic experimental conditions.

Finally, we make a comparison between the pulse modulation strategy for gases and solids. The regulation of pulse generation in solids should qualitatively resemble that in gases; however, the diverse atomic and electronic structures of the former lead to intricate carrier dynamics. Consequently, the quantitative dependence of harmonic properties on laser parameters, such as the relationship between the cutoff energy and laser frequency, exhibit significant differences. For instance, in gases, $\varepsilon_c=3.17\frac{e^{2} E_{p}^{2}}{4m\omega^{2}}+I_{p}$, which is inversely proportional to the square of the laser frequency \cite{ghimire2019high}. In solids, considering the field-dressed energy gap in adiabatic dynamics, $\varepsilon_c\propto \frac{E_{p}}{\omega } $ in our designed optical field (See Appendix \ref{app:cutoff}), which is approximately inversely proportional to the laser frequency \cite{wu2016multilevel}. The differences of harmonic properties between solids and gases may lead to quantitative differences in pulse generation.

\section{SUMMARY}
In conclusion, we have demonstrated an effective approach for generating attosecond pulses by manipulating the waveforms of monochromatic laser pulses. This approach aims to monotonically increase the peak value of the vector potential by modifying the parameters of the optical field. Our analysis points out that increasing the vector potential enhances both interband and intraband dynamics of electrons, which leads to more electrons moving to higher energy conduction bands. This, in turn, helps to strengthen the intensity of the highest plateau region near the cutoff, resulting in pulses with narrower width and higher intensity obtained from this region. Although the exact main pulse emission time can occur at either the peak or the secondary peak of the vector potential in different situations, our optical field modulation approach can still monotonically increase the secondary peak of the vector potential over a broad frequency range. However, at low frequencies, the broaden of harmonic emission time range becomes dominate. Therefore, this inevitably leads to non-monotonic behavior in both pulse width and intensity. By carefully adjusting the optical field parameters, we can obtain the optimal pulse. 
Our modulation scheme significantly reduces pulse width and enhances pulse intensity in 1L-hBN theoretically compared to other strategies. Preliminary simulations on other solids further illustrate the broad applicability of our approach for optimal attosecond pulse generation. However, there are still many experimental challenges, such as propagation-induced atto-chirps, precise laser parameter control, and attosecond pulse measurement, are critical for practical applications.

\textbf{ACKNOWLEDGEMENT} 
This work is supported by the National Natural Science Foundation of China (Grants No. 12450401, No. 12025407,
No. 12304536, and No. 92250303), the Ministry of Science and Technology (Grant No. 2021YFA1400201), and the Chinese Academy of Sciences (Grants No. YSBR047 and No. XDB33030100). M.-X.G. acknowledges support from the start-up funding of Beijing Institute of Technology.

\appendix

\section{BAND STRUCTURE OF 1L-$\text{hBN}$}

Compared to bulk materials, two-dimensional monolayers demonstrate a higher efficiency in generating high-order harmonics due to weaker screening effects \cite{tancogne2017impact,liu2017high}. With its wide direct gap, 1L-hBN is less susceptible to damage under intense light and likely possesses a large cutoff energy, making it ideal for solid-state AP generation \cite{elias2019direct}.

\begin{figure}[!htbp]
\centering
\includegraphics[width=8.6cm]{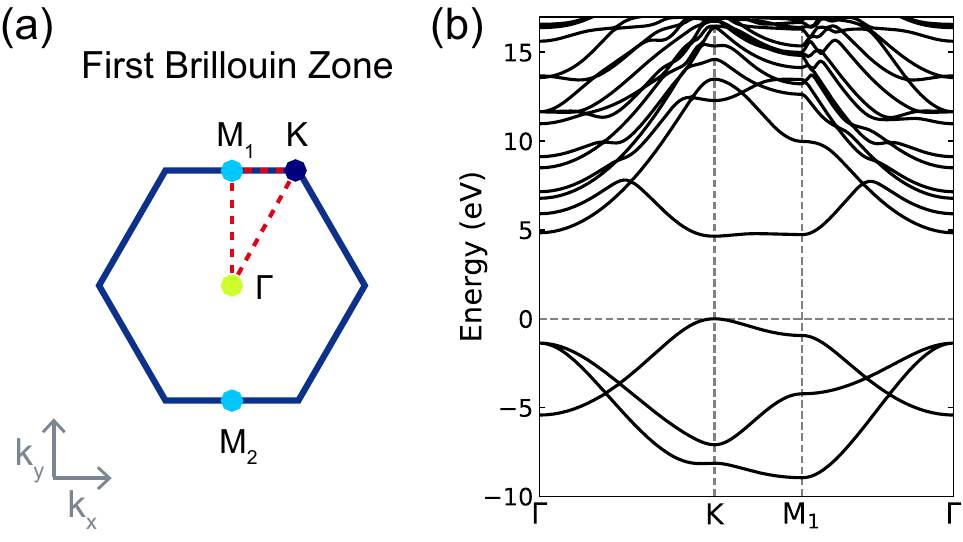}
\caption{ (a) The FBZ of 1L-hBN. (b) Band structure of 1L-hBN. }\label{fig;bandstructure}
\end{figure}

The ground state properties of 1L-hBN are calculated using density functional theory (DFT). The first Brillouin zone (FBZ) and band structure of 1L-hBN are shown in Fig. \ref{fig;bandstructure}. A direct gap of 4.6 eV is found at the K point, attributed to the underestimation of the gap value by the LDA functional.

\section{THEORETICAL BACKGROUND} \label{app:TDDFT}
Time-dependent density functional theory (TDDFT) describes laser-matter interactions during harmonic generation. The evolution of electron wavefunctions is calculated using the time-dependent Kohn-Sham (TDKS) equation in the velocity gauge \cite{ullrich2011time},
\begin{align}
i\frac{\partial}{\partial t}\phi_{m,\mathbf{k}}\left( \mathbf{r} ,t \right)=\left( \frac{1}{2}\left( -i\nabla+\mathbf{A} \left( t \right) \right)^{2}+v_{KS}\left( \mathbf{r},t \right) \right)\phi_{m,\mathbf{k}}\left( \mathbf{r} ,t \right),
\end{align}
where $\phi_{m,\mathbf{k}}\left( \mathbf{r} ,t \right)$ represents the time-dependent Kohn-Sham wavefunction and the subscripts $m$ and $\textbf{k}$ denote the band and \textbf{\textit{k}}-point indexes. $\mathbf{A} \left( t \right)$ corresponds to the vector potential of the laser within the dipole approximation, and \textit{c} represents the speed of light. The current density is given by 
\begin{align}
\mathbf{j}\left( \mathbf{r} ,t \right)=\sum_{m,\mathbf{k}}^{}Re\left[ \phi_{m,\mathbf{k}}^{\ast}\left( \mathbf{r} ,t \right) \left( -i\nabla +\mathbf{A}\left( t \right)  \right)\phi_{m,\mathbf{k}}\left( \mathbf{r} ,t \right) \right].
\end{align}

The high harmonic generation (HHG) intensity can be calculated by applying the Fourier transform (FT) to the dipole acceleration
\begin{align}
\mathrm{HHG}\left( \omega \right) =\left| \text{FT} \left[ \frac{\partial}{\partial t}\int_{\Omega }d^{3}\mathbf{r} \mathbf{j}\left( \mathbf{r} ,t \right) \right] \right|^{2},
\end{align}
here, the system volume is $\Omega$. The amplitude gating method, using a filter to select a specific frequency range of high harmonics, is used to obtain ultrashort pulse signals in the time domain. The attosecond pulse is then generated by applying the inverse Fourier transform  
\begin{align}
I\left( t \right) =\left| \sum_{\omega _{i} }^{\omega _{f}}e^{i\omega t}\text{FT}\left[ \frac{\partial}{\partial t}\int_{\Omega }d^{3}\mathbf{r} \mathbf{j}\left( \mathbf{r} ,t \right) \right] \right|^{2},
\end{align}
$\hbar\omega _{i}$ and $\hbar\omega _{f}$ refer to the minimum and maximum values of the selected energy window, respectively.

To further investigate the electron dynamics during the process of HHG, the electronic occupancy in conduction bands is calculated
\begin{align}
\Delta n_{m}\left( \mathbf{k},t \right)=\frac{1}{N}\sum_{l}^{occ} \left| \left \langle \phi_{l,\mathbf{k}}\left(  \mathbf{r},t \right)  | \phi_{m,\mathbf{k}}\left( \mathbf{r},0 \right)  \right \rangle \right|^{2},
\end{align}
where $N$ is the total number of \textbf{\textit{k}}-points. Then by summing the number of excited electrons of the considered \textbf{\textit{k}}-points, we can obtain the total excited number in $m^{th}$ conduction bands along the high-symmetry line in the FBZ within the primitive cell.

In our work, we used the OCTOPUS package to perform TDDFT simulations \cite{tancogne2020octopus}. The 1L-hBN model has a 2.48 $\AA$ in-plane lattice constant and a 0.16 $\AA$ grid spacing. A $21\times21\times1$ \textbf{\textit{k}}-point grid centered around the $\Gamma$ point samples the reciprocal space, and Troullier-Martins pseudopotentials within the adiabatic local-density approximation describe electron-ion interactions during HHG.

\section{DETAILS OF THE FOUR-LEVEL MODEL} \label{app:four}

\begin{table}[!htbp] 
\centering
\begin{ruledtabular}
\begin{tabular}{ccccccc}
Level&2$^{nd}$ VB&1$^{st}$ VB&8$^{th}$ CB&9$^{th}$ CB&10$^{th}$ CB&Eigenvalue (a.u.) \\
2$^{nd}$ VB&0&0&0.57&0.40&0&-0.050 \\
1$^{st}$ VB&0&0&0.40&0.57&0&-0.050 \\
8$^{th}$ CB&0.57&0.40&0&0&0&0.428 \\
9$^{th}$ CB&0.40&0.57&0&0&0&0.428 \\ 
10$^{th}$ CB&0&0&0&0&0&0.428 \\ 
\end{tabular}
\end{ruledtabular}
\caption{Transition matrix elements among 1$^{st}$-2$^{nd}$ VBs, 8$^{th}$-10$^{th}$ CBs at $\Gamma$ point. The last column means the eigenvalue of corresponding state.  All physical quantities are given in atomic units (a.u.).}
\label{tab;transition matrix}
\end{table}

The transition matrix elements among the five states (1$^{st}$-2$^{nd}$ VBs, 8$^{th}$-10$^{th}$ CBs) are detailed in Table \ref{tab;transition matrix}. Since the transition matrix elements involving the 10$^{th}$ CB are zero, our model is limited to the Hamiltonian within the subspace of states that include the 1$^{st}$-2$^{nd}$ VBs and the 8$^{th}$-9$^{th}$ CBs at the $\Gamma$ point.

The time-dependent hamiltonian, formulated in atomic units, is expressed as:
\begin{align}
H\left( t \right)= \frac{\left( \mathbf{p}+\mathbf{A}\left( t \right) \right)^{2}  }{2}+V\left ( r \right ),
\end{align} 
where $V\left ( r \right )$ denotes the single-electron potential. In the presence of an external field, the Hamiltonian matrix can be diagonalized at each time step. The Hamiltonian matrix within the four-level subspace is formulated as follows:
\begin{align}
H=\begin{bmatrix}
  \omega_{2}+\frac{A\left( t \right)^{2} }{2}& \mu_{21}A\left( t \right)& \mu_{28}A\left( t \right)&\mu_{29}A\left( t \right) \\
  \mu_{12}A\left( t \right)&\omega_{1}+\frac{A\left( t \right)^{2} }{2}& \mu_{18}A\left( t \right) &\mu_{19}A\left( t \right)  & \\
  \mu_{82}A\left( t \right)& \mu_{81}A\left( t \right) & \omega_{8}+\frac{A\left( t \right)^{2} }{2} & \mu_{89}A\left( t \right) & \\
  \mu_{92}A\left( t \right)& \mu_{91}A\left( t \right) & \mu_{98}A\left( t \right) & \omega_{9}+\frac{A\left( t \right)^{2} }{2} &
\end{bmatrix}
\end{align}
Here, the subscript denotes the energy level. $\omega_n$ represents the eigenvalue of the state and $\mu_{mn}$ represents the transition matrix element: $\mu_{mn}=\left \langle \phi_{m}  \right| \mathbf{p} \left | \phi_{n}  \right \rangle _{cell}$. This allows for the calculation of time-dependent eigenvalues of adiabatic states. The energy differences between these states predict the potential harmonic energies generated by electron recombination from CBs to VBs, with the maximum value being $\varepsilon_c$. Thus, we can determine the dependence of $\varepsilon_c$ as a function of phonon energy $\hbar\omega$.

\section{THE DEPENDENCE OF THE TOTAL CUTOFF ENERGY ON LASER FREQUENCY} \label{app:cutoff}
\begin{figure}[!htbp]
\centering
\includegraphics[width=0.3\textwidth]{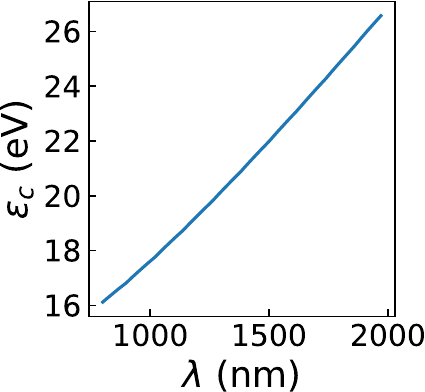}
\caption{Numerical calculation of the total cutoff energy $\varepsilon_c$ of different field wavelength ($\lambda$) in our four-level model. }\label{fig;linear}
\end{figure}

Since the four-level model in our work can not provide an analytical expression for the cutoff energy, we first present the results from the two-level model in ref. \cite{wu2016multilevel}:
\begin{align}
\varepsilon_{c}=2\sqrt{\left ( \mu A_{p} \right )^{2} +  \left (  \frac{\omega_{0}}{2}\right )^{2}  },
\label{cutoff}
\end{align} 
where $\omega_{0}$ and $\mu$ are the energy gap and transition matrix respectively. Under our field configuration, the vector potential is given by:
\begin{align}
A\left ( t \right ) =-\int_{0}^{t}E_{p} g\left ( t \right )\cos\left( \omega \left ( t-t_{0}  \right ) \right)dt,
\end{align}
and since the Gaussian envelope varies slowly, it can be neglected. The peak vector potential can then be approximated as:
\begin{align}
A_{p} \approx \frac{1}{\omega}E_{p}.
\end{align}
Furthermore, at high field strengths, the second term in Eq. \ref{cutoff} becomes negligible, yielding:	
\begin{align}
\varepsilon_{c} \approx 2\sqrt{\left ( \mu A_{p} \right )^{2}} \approx \frac{2\mu}{\omega}E_{p},
\end{align}
thus, the cutoff energy is linearly proportional to the inverse of $\omega$, meaning that the cutoff energy scales linearly with the wavelength $\lambda$. Although this analysis relies on the two-level model \cite{wu2016multilevel}, our system's four-level model yields the same conclusion. Fig. \ref{fig;linear} shows numerical calculations using our four-level model, confirming this linear relationship without applying any approximations. 

\section{TIME DEPENDENT PERTURBATION THEORY AND TRANSITION PROBABILITY} \label{app:tdpt}
The single electron Hamiltonian under the velocity gauge and dipole approximation is written as
\begin{align}
H= H_{0}+ H'\left( t \right) =H_{0}+\mathbf{A}\left( t \right)\cdot \mathbf{p}+ \frac{1}{2}\mathbf{A\left( t \right)^{2}},
\end{align}
where $H_{0}$ is the ground-state Hamiltonian of the system. The time-evolving wavefunction is expanded in Bloch basis $\left| \psi \left( t \right) \right\rangle=\sum_{m}\sum_{\mathbf{q}\in BZ}b_{m}^{\mathbf{q}}\left( t \right)e^{-iE_{m}^{\mathbf{q}}t}\left| \phi_{m}^{\mathbf{q}} \right\rangle$. Substitute this expansion into the time-dependent Schrödinger equation and multiply $\left\langle \phi_{n}^{k} \right| $ from the left to acquire inner product in the crystal. The evolution equation of the expansion coefficients is expressed by
\begin{align} 
\frac{\partial}{\partial t}b_{n}^{\mathbf{k}}\left( t \right)=-\frac{i}{N}\sum_{m}\sum_{\mathbf{q}\in BZ}\left\langle \phi_{n}^{\mathbf{k}} \right|H'\left( t \right)\left| \phi_{m}^{\mathbf{q}} \right\rangle_{crys}b_{m}^{\mathbf{q}}\left( t \right) e^{i\left( E_{n}^{\mathbf{k}}-E_{m}^{\mathbf{q}} \right) t} ,
\label{bnk}
\end{align}
where $N$ is number of primitive cells. Neglecting the second-order term of $\mathbf{A}$, the matrix element in Eq. (\ref{bnk}) is given by $\left\langle \phi_{n}^{\mathbf{k}} \right|\mathbf{A}\left ( t \right )\cdot \mathbf{p} \left| \phi_{m}^{\mathbf{q}} \right\rangle_{crys}$. Due to the orthogonality of $\mathbf{p}$ matrix in the Bloch basis $\left\langle \phi_{n}^{\mathbf{k}} \right| \mathbf{p} \left| \phi_{m}^{\mathbf{q}} \right\rangle_{crys}=N\delta_{\mathbf{k},\mathbf{q}}\left\langle \phi_{n}^{\mathbf{k}} \right| \mathbf{p} \left| \phi_{m}^{\mathbf{q}} \right\rangle_{cell}$, we obtain the simplified equation of $b_{n}^{\mathbf{k}}\left( t \right)$
\begin{align} 
\frac{\partial}{\partial t}b_{n}^{\mathbf{k}}\left( t \right)=-i\sum_{m} \mathbf{A}\left ( t \right )\cdot \left\langle \phi_{n}^{\mathbf{k}} \right| \mathbf{p} \left| \phi_{m}^{\mathbf{k}} \right\rangle_{cell}b_{m}^{\mathbf{k}}\left( t \right) e^{i\left( E_{n}^{\mathbf{k}}-E_{m}^{\mathbf{k}} \right) t} .
\label{bnks}
\end{align}

Now we use time-dependent perturbation theory to calculate Eq. (\ref{bnks}). Assuming that the initial electron state is $\left| \phi_{n_{0}}^{\mathbf{k}} \right\rangle$, the zeroth-order approximation of the expansion coefficients can be represented as $b_{n}^{\mathbf{k}}=\delta_{n,n_{0}}$. Then substituting the zeroth-order approximation into Eq. (\ref{bnks}) and integrating over time, the first-order approximation of $b_{n}^{\mathbf{k}} \left( n\ne n_{0} \right) $ satisfies the equation
\begin{align} 
b_{n}^{\mathbf{k}}\left( t \right)=-i \left\langle \phi_{n}^{\mathbf{k}} \right| \mathbf{p} \left| \phi_{n_{0}}^{\mathbf{k}} \right\rangle_{cell} \cdot \int_{0}^{t} \mathbf{A}\left ( t \right ) e^{i\left( E_{n}^{\mathbf{k}}-E_{n_{0}}^{\mathbf{k}} \right) t} dt,
\end{align}
at time $t$, the number of electrons in state $\left| \phi_{n}^{\mathbf{k}} \right\rangle$ excited from $\left| \phi_{n_{0}}^{\mathbf{k}} \right\rangle$ is
\begin{align}
\left| b_{n}^{\mathbf{k}}\left( t \right) \right|^{2} = \left|\left\langle \phi_{n}^{\mathbf{k}} \right| \mathbf{p} \left| \phi_{n_{0}}^{\mathbf{k}} \right\rangle_{cell} \cdot \int_{0}^{t} \mathbf{A}\left ( t \right )  e^{i\left( E_{n}^{\mathbf{k}}-E_{n_{0}}^{\mathbf{k}} \right) t} dt \right |^{2} .
\label{r}
\end{align} 

Taking into account the relation $\mathbf{d}_{nn_{0}}^{\mathbf{k}}=i\left\langle u_{n}^{\mathbf{k}} \right| \nabla_\mathbf{k} \left| u_{n_{0}}^{\mathbf{k}} \right\rangle_{cell}=\frac{-i}{\left( E_{n}^{\mathbf{k}}-E_{n_{0}}^{\mathbf{k}} \right)} \left\langle \phi_{n}^{\mathbf{k}} \right| \mathbf{p} \left| \phi_{n_{0}}^{\mathbf{k}} \right\rangle_{cell}$, $u_{n}^{\mathbf{k}}=e^{-i\mathbf{k}\cdot \mathbf{r}} \phi_{n}^{\mathbf{k}}$, Eq. (\ref{r}) is rewritten as
\begin{align}
\left| b_{n}^{\mathbf{k}}\left( t \right) \right|^{2} = \left( E_{n}^{\mathbf{k}}-E_{n_{0}}^{\mathbf{k}} \right)^2 \left|\left\langle u_{n}^{\mathbf{k}} \right| \nabla_\mathbf{\mathbf{k}} \left| u_{n_{0}}^{\mathbf{k}} \right\rangle_{cell} \cdot \int_{0}^{t} \mathbf{A}\left ( t \right )  e^{i\left( E_{n}^{\mathbf{k}}-E_{n_{0}}^{\mathbf{k}} \right) t} dt \right|^{2},
\label{b}
\end{align}
\begin{align}
f_{n_{0}\to n}^{\mathbf{k}} \left (t\right )=\left |\int_{0}^{t} A\left(t\right) e^{\frac{i\left(E_{n}^{\mathbf{k}}-E_{n_{0}}^{\mathbf{k}}\right)t}{\hbar}} dt \right|^{2},
\label{f}
\end{align}
$f_{n_{0}\to n}^{\mathbf{k}} \left (t\right )$ is the time-dependent function in Eq. (\ref{b}) related to the vector potential.

Considering the time reversal symmetry of the unperturbed Hamiltonian $H_{0}$, the ground-state wavefunctions satisfy ${u_{n}^{-\mathbf{k}}}^{\ast } =u_{n}^{\mathbf{k}}$, $E_{n}^{-\mathbf{k}} =E_{n}^{\mathbf{k}}$ and ${\mathbf{d}_{nn_{0}}^{-\mathbf{k}}}^{\ast }=\mathbf{d}_{nn_{0}}^{\mathbf{k}}$. Based on the above relations, the number of electrons in the $n^{th}$ band at the \textit{-\textbf{k}}-point is the same as at the \textit{\textbf{k}}-point along the field direction
\begin{align}
\left| b_{n}^{-\mathbf{k}}\left( t \right) \right|^{2}=\left| b_{n}^{\mathbf{k}}\left( t \right) \right|^{2},
\end{align}
this indicates that the electron distribution in reciprocal space will remain unchanged, i.e., intraband electronic motion is not involved in this theory.

\begin{figure}[!htbp]
\centering
\includegraphics[width=8.6cm]{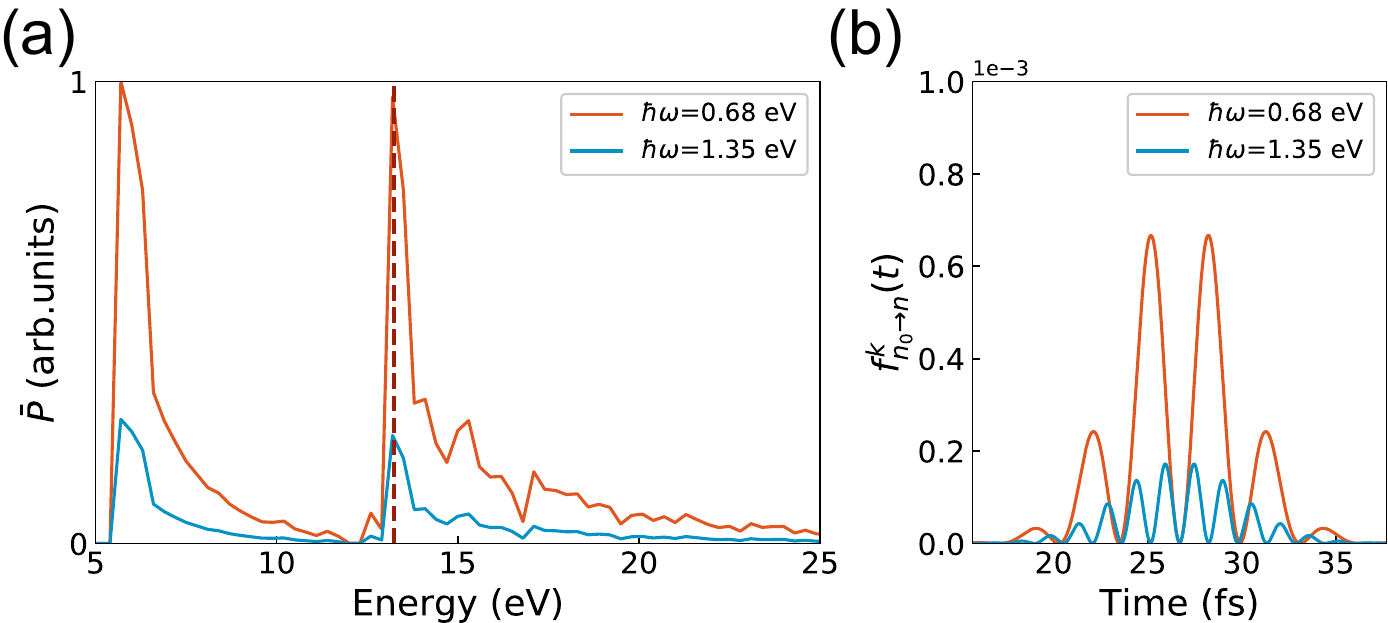}
\caption{(a) The overall transition probability, denoted as $\bar{P}$, as functions of photon energy. The vertical dark red dashed line is at 12.9 eV. (b) The waveform of $f_{n_{0}\to n}^{\mathbf{k}} \left (t\right )$ from the 1$^{st}$ VB to the 8$^{th}$ CB of $\Gamma$ point at photon energies of 0.68 eV and 1.35 eV. }\label{fig;tdpt}
\end{figure}

For 1L-hBN, according to Eq. (\ref{b}), the electronic transition probability from band $n_{0}$ to band $n$ is written as:
\begin{align}
P_{n_{0}\to n}^{\mathbf{k}}\left (t\right )= \left| b_{n}^{\mathbf{k}}\left( t \right) \right|^{2}.
\label{p}
\end{align}

The time average of electronic transition probability is calculated by
\begin{align}
\bar{P}_{n_{0}\to n}^{\mathbf{k}}=\frac{1}{T_o}\int_{0}^{T_o} P_{n_{0}\to n}^{\mathbf{k}} dt,
\label{p1}
\end{align}
$T_o$ is the optical field duration.

For a given \textbf{\textit{k}}-point in FBZ, we can calculate the energy difference $E_{n}^{\mathbf{k}}-E_{n_{0}}^{\mathbf{k}}$ for each electronic transition. Subsequently, the transition probability $\bar{P}_{\mathbf{k}}$ is shown as a function of energy. The overall transition probability $\bar{P}$ along the high-symmetry line can be obtained by performing an average of $\bar{P}_{\mathbf{k}}$ over all $\textit{\textbf{k}}$-points (Fig. \ref{fig;tdpt}(a)). A step-like structure appears in $\bar{P}$, with the second peak precisely at 12.9 eV, indicated by the vertical dark red dashed line. Beyond this value, direct transitions of electrons with high energy differences contribute to the high-energy region of the harmonic spectrum.

As for the time-dependent part $f_{n_{0}\to n}^{\mathbf{k}} \left (t\right )$ of Eq. (\ref{f}), since the photon energy (0.68-1.55 eV) is much lower than the minimum transition energy gap (4.6 eV in our case), the external field $\mathbf{A}\left ( t \right )$ serves as the slowly varying envelope part of the integrand, while the exponential term determines the rapidly oscillating carrier part. We apply integration by parts:
\begin{multline}
\int_{0}^{t} \mathbf{A}(t) e^{i\left(E_{n}^{\mathbf{k}}-E_{n_{0}}^{\mathbf{k}}\right) t} d t=\frac{\mathbf{A}(t)}{i\left(E_{n}^{\mathbf{k}}-E_{n_{0}}^{\mathbf{k}}\right)} e^{i\left(E_{n}^{\mathbf{k}}-E_{n_{0}}^{\mathbf{k}}\right) t}\\-\frac{1}{i\left(E_{n}^{\mathbf{k}}-E_{n_{0}}^{\mathbf{k}}\right)} \int_{0}^{t} \dot{\mathbf{A}}(t) e^{i\left(E_{n}^{\mathbf{k}}-E_{n_{0}}^{\mathbf{k}}\right) t} d t, 
\label{tintegral}
\end{multline}

\begin{multline}
\text{Second term of Eq. (\ref{tintegral})}= \frac{\dot{\mathbf{A}}\left ( t \right )}{\left(E_{n}^{\mathbf{k}}-E_{n_{0}}^{\mathbf{k}}\right)^{2}}e^{i\left(E_{n}^{\mathbf{k}}-E_{n_{0}}^{\mathbf{k}}\right)t}\\-\frac{1}{\left(E_{n}^{\mathbf{k}}-E_{n_{0}}^{\mathbf{k}}\right)^{2}}\int_{0}^{t} \ddot{\mathbf{A}} \left ( t \right ) e^{i\left(E_{n}^{\mathbf{k}}-E_{n_{0}}^{\mathbf{k}}\right)t}dt.
\end{multline}
The second term is small compared to the first term because the factor $\frac{1}{\left(E_{n}^{\mathbf{k}}-E_{n_{0}}^{\mathbf{k}}\right)^{2}}$ represents the higher-order contributions, and $\dot{\mathbf{A}}\left ( t \right )$ is small due to the slowly varying envelope part. $f_{n_{0}\to n}^{\mathbf{k}} \left (t\right )$ can thus be approximately written as:
\begin{align}
f_{n_{0}\to n}^{\mathbf{k}} \left (t\right )\approx \left | \frac{\mathbf{A}\left ( t \right )}{i\left(E_{n}^{\mathbf{k}}-E_{n_{0}}^{\mathbf{k}}\right)} e^{i\left(E_{n}^{\mathbf{k}}-E_{n_{0}}^{\mathbf{k}}\right) t}  \right |^{2}=\frac{\mathbf{A}\left ( t \right )^{2} }{\left(E_{n}^{\mathbf{k}}-E_{n_{0}}^{\mathbf{k}}\right)^{2}}.   
\end{align}
Therefore, the maximum value of $f_{n_{0}\to n}^{\mathbf{k}} \left (t\right )$ is still determined by $A_p$, and the moments when $\left | \mathbf{A}\left ( t \right ) \right | $ and $f_{n_{0}\to n}^{\mathbf{k}} \left (t\right )$ reach their extrema coincide.

\begin{figure}[!htbp]
\centering
\includegraphics[width=8.6cm]{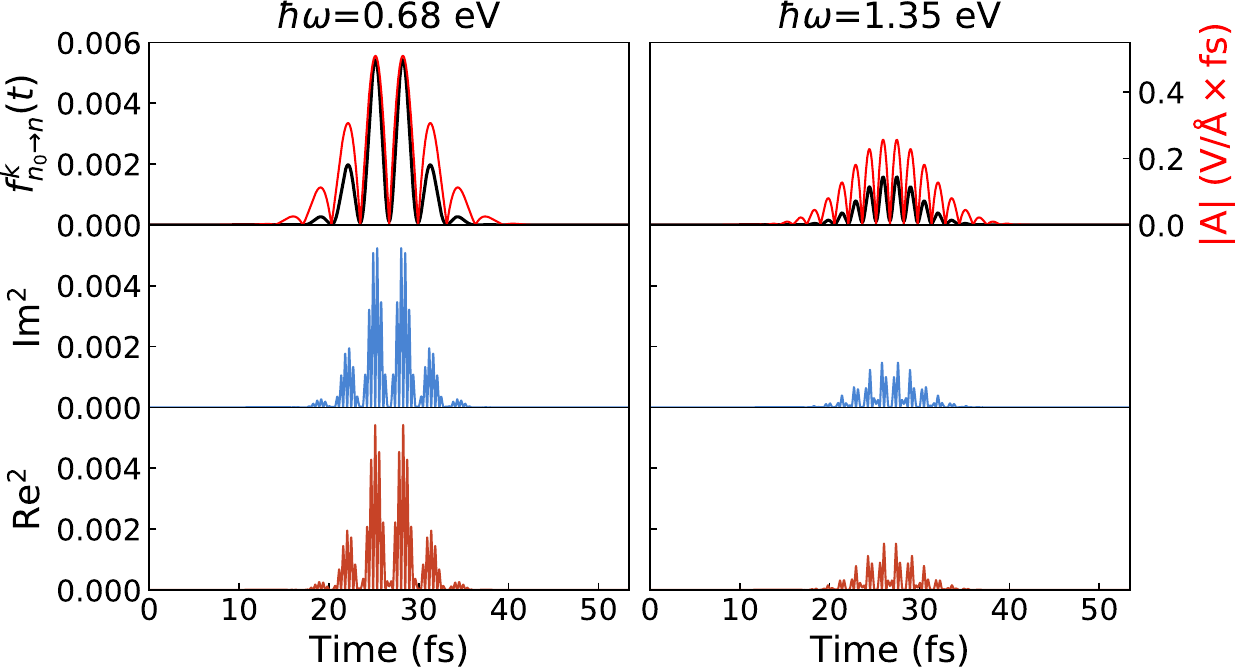}
\caption{(a) Calculation of the transition at the energy gap for different photon energies, showing $f_{n_{0}\to n}^{\mathbf{k}} \left (t\right )$ and $\left | \mathbf{A}\left ( t \right ) \right | $ in the upper graphs, and the square of real and imaginary parts of the time integral in the middle and lower graphs.}\label{fig;gaptransition}
\end{figure}

Furthermore, based on Eq. (\ref{f}) and without applying any of the above approximations, Fig. \ref{fig;gaptransition} shows the numerical calculation of the square of real and imaginary parts of the time integral as well as a comparison of $f_{n_{0}\to n}^{\mathbf{k}} \left (t\right )$ with $\left | \mathbf{A}\left ( t \right ) \right | $ at photon energies of 0.68 and 1.35 eV. As can be seen, the extrema of the vector potential coincide with those of $f_{n_{0}\to n}^{\mathbf{k}} \left (t\right )$. As $A_p$ increases, the maximum value of $f_{n_{0}\to n}^{\mathbf{k}} \left (t\right )$ also increases. These results, consistent with the above approximations, demonstrate the validity of our approximations.

Considering the high-energy excitations near the $\Gamma$ point that are of interest in the main text, we also performed numerical calculations for those transitions (e.g., from the 1$^{st}$ VB to the 8$^{th}$ CB). At 0.68 eV, $f_{n_{0}\to n}^{\mathbf{k}} \left (t\right )$ exhibits higher peaks compared to 1.35 eV (Fig. \ref{fig;tdpt}(b)). Consequently, $P_{n_{0}\to n}^{\mathbf{k}}\left (t\right )$ and the total number of excited electrons are also larger at 0.68 eV (see Fig. \ref{fig;direct excitation}(c) in the main text).

\section{PULSE MODULATION IN MONOLAYER $\text{GaSe}$ and Wurtzite-structured $\text{ZnO}$} \label{app:other}
\begin{figure}[!htbp]
\centering
\includegraphics[width=8.6cm]{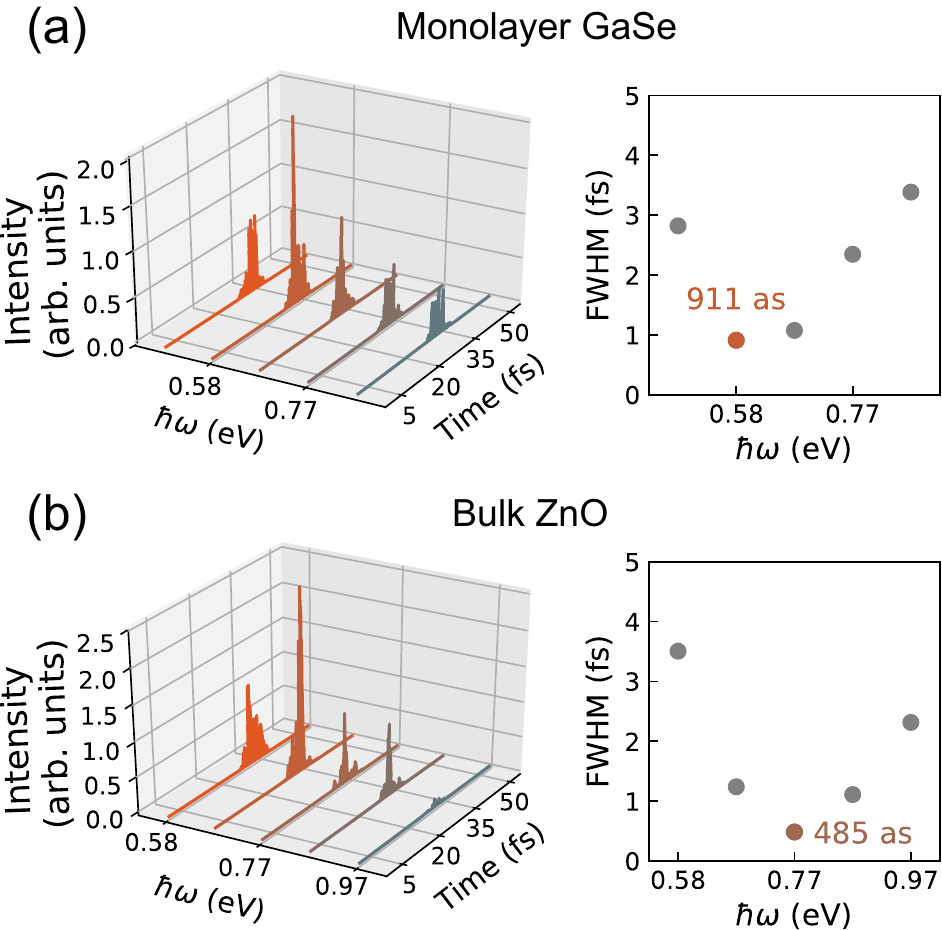}
\caption{(a) Time-domain ultrashort pulses filtered from the high harmonic spectra of monolayer GaSe within the 32–60 eV energy range at different photon energies, with the FWHM of the pulses plotted against photon energy. (b) Time-domain ultrashort pulses filtered from the high harmonic spectra of bulk ZnO within the 16–60 eV energy range at different photon energies, with the FWHM of the pulses plotted against photon energy.}\label{fig;gase and zno}
\end{figure}

The modulation methods discussed in the main text are applicable to other materials, including monolayer GaSe and bulk ZnO, which exhibit significant potential as nonlinear optical materials for HHG. In this context, we conducted preliminary tests on these materials at various frequencies. In monolayer GaSe, as depicted in Fig. \ref{fig;gase and zno}(a), decreasing the laser frequency initially reduces the pulse's FWHM, followed by an increase, while the pulse intensity first increases and then decreases. This behavior is similar to what is observed in 1L-hBN. Notably, the narrowest pulse occurs at $\hbar\omega = 0.58$ eV, with a FWHM of 911 as. Similarly, as depicted in Figs. \ref{fig;gase and zno}(b), the properties of generated ultrashort pulse in bulk ZnO exhibit the same trend, with the narrowest pulse at $\hbar\omega = 0.77$ eV, achieving a FWHM of 485 as.

\bibliography{main.bbl}

\begin{thebibliography}{50}%
\makeatletter
\providecommand \@ifxundefined [1]{%
 \@ifx{#1\undefined}
}%
\providecommand \@ifnum [1]{%
 \ifnum #1\expandafter \@firstoftwo
 \else \expandafter \@secondoftwo
 \fi
}%
\providecommand \@ifx [1]{%
 \ifx #1\expandafter \@firstoftwo
 \else \expandafter \@secondoftwo
 \fi
}%
\providecommand \natexlab [1]{#1}%
\providecommand \enquote  [1]{``#1''}%
\providecommand \bibnamefont  [1]{#1}%
\providecommand \bibfnamefont [1]{#1}%
\providecommand \citenamefont [1]{#1}%
\providecommand \href@noop [0]{\@secondoftwo}%
\providecommand \href [0]{\begingroup \@sanitize@url \@href}%
\providecommand \@href[1]{\@@startlink{#1}\@@href}%
\providecommand \@@href[1]{\endgroup#1\@@endlink}%
\providecommand \@sanitize@url [0]{\catcode `\\12\catcode `\$12\catcode `\&12\catcode `\#12\catcode `\^12\catcode `\_12\catcode `\%12\relax}%
\providecommand \@@startlink[1]{}%
\providecommand \@@endlink[0]{}%
\providecommand \url  [0]{\begingroup\@sanitize@url \@url }%
\providecommand \@url [1]{\endgroup\@href {#1}{\urlprefix }}%
\providecommand \urlprefix  [0]{URL }%
\providecommand \Eprint [0]{\href }%
\providecommand \doibase [0]{https://doi.org/}%
\providecommand \selectlanguage [0]{\@gobble}%
\providecommand \bibinfo  [0]{\@secondoftwo}%
\providecommand \bibfield  [0]{\@secondoftwo}%
\providecommand \translation [1]{[#1]}%
\providecommand \BibitemOpen [0]{}%
\providecommand \bibitemStop [0]{}%
\providecommand \bibitemNoStop [0]{.\EOS\space}%
\providecommand \EOS [0]{\spacefactor3000\relax}%
\providecommand \BibitemShut  [1]{\csname bibitem#1\endcsname}%
\let\auto@bib@innerbib\@empty
\bibitem [{\citenamefont {Mashiko}\ \emph {et~al.}(2016)\citenamefont {Mashiko}, \citenamefont {Oguri}, \citenamefont {Yamaguchi}, \citenamefont {Suda},\ and\ \citenamefont {Gotoh}}]{mashiko2016petahertz}%
  \BibitemOpen
  \bibfield  {author} {\bibinfo {author} {\bibfnamefont {H.}~\bibnamefont {Mashiko}}, \bibinfo {author} {\bibfnamefont {K.}~\bibnamefont {Oguri}}, \bibinfo {author} {\bibfnamefont {T.}~\bibnamefont {Yamaguchi}}, \bibinfo {author} {\bibfnamefont {A.}~\bibnamefont {Suda}},\ and\ \bibinfo {author} {\bibfnamefont {H.}~\bibnamefont {Gotoh}},\ }\bibfield  {title} {\bibinfo {title} {Petahertz optical drive with wide-bandgap semiconductor},\ }\href {https://www.nature.com/articles/nphys3711} {\bibfield  {journal} {\bibinfo  {journal} {Nature physics}\ }\textbf {\bibinfo {volume} {12}},\ \bibinfo {pages} {741} (\bibinfo {year} {2016})}\BibitemShut {NoStop}%
\bibitem [{\citenamefont {Sch{\"o}tz}\ \emph {et~al.}(2019)\citenamefont {Sch{\"o}tz}, \citenamefont {Wang}, \citenamefont {Pisanty}, \citenamefont {Lewenstein}, \citenamefont {Kling},\ and\ \citenamefont {Ciappina}}]{schotz2019perspective}%
  \BibitemOpen
  \bibfield  {author} {\bibinfo {author} {\bibfnamefont {J.}~\bibnamefont {Sch{\"o}tz}}, \bibinfo {author} {\bibfnamefont {Z.}~\bibnamefont {Wang}}, \bibinfo {author} {\bibfnamefont {E.}~\bibnamefont {Pisanty}}, \bibinfo {author} {\bibfnamefont {M.}~\bibnamefont {Lewenstein}}, \bibinfo {author} {\bibfnamefont {M.~F.}\ \bibnamefont {Kling}},\ and\ \bibinfo {author} {\bibfnamefont {M.}~\bibnamefont {Ciappina}},\ }\bibfield  {title} {\bibinfo {title} {Perspective on petahertz electronics and attosecond nanoscopy},\ }\href {https://pubs.acs.org/doi/full/10.1021/acsphotonics.9b01188} {\bibfield  {journal} {\bibinfo  {journal} {ACS photonics}\ }\textbf {\bibinfo {volume} {6}},\ \bibinfo {pages} {3057} (\bibinfo {year} {2019})}\BibitemShut {NoStop}%
\bibitem [{\citenamefont {F{\"o}hlisch}\ \emph {et~al.}(2005)\citenamefont {F{\"o}hlisch}, \citenamefont {Feulner}, \citenamefont {Hennies}, \citenamefont {Fink}, \citenamefont {Menzel}, \citenamefont {S{\'a}nchez-Portal}, \citenamefont {Echenique},\ and\ \citenamefont {Wurth}}]{fohlisch2005direct}%
  \BibitemOpen
  \bibfield  {author} {\bibinfo {author} {\bibfnamefont {A.}~\bibnamefont {F{\"o}hlisch}}, \bibinfo {author} {\bibfnamefont {P.}~\bibnamefont {Feulner}}, \bibinfo {author} {\bibfnamefont {F.}~\bibnamefont {Hennies}}, \bibinfo {author} {\bibfnamefont {A.}~\bibnamefont {Fink}}, \bibinfo {author} {\bibfnamefont {D.}~\bibnamefont {Menzel}}, \bibinfo {author} {\bibfnamefont {D.}~\bibnamefont {S{\'a}nchez-Portal}}, \bibinfo {author} {\bibfnamefont {P.~M.}\ \bibnamefont {Echenique}},\ and\ \bibinfo {author} {\bibfnamefont {W.}~\bibnamefont {Wurth}},\ }\bibfield  {title} {\bibinfo {title} {Direct observation of electron dynamics in the attosecond domain},\ }\href {https://www.nature.com/articles/nature03833} {\bibfield  {journal} {\bibinfo  {journal} {nature}\ }\textbf {\bibinfo {volume} {436}},\ \bibinfo {pages} {373} (\bibinfo {year} {2005})}\BibitemShut {NoStop}%
\bibitem [{\citenamefont {Schultze}\ \emph {et~al.}(2010)\citenamefont {Schultze}, \citenamefont {Fie{\ss}}, \citenamefont {Karpowicz}, \citenamefont {Gagnon}, \citenamefont {Korbman}, \citenamefont {Hofstetter}, \citenamefont {Neppl}, \citenamefont {Cavalieri}, \citenamefont {Komninos}, \citenamefont {Mercouris} \emph {et~al.}}]{schultze2010delay}%
  \BibitemOpen
  \bibfield  {author} {\bibinfo {author} {\bibfnamefont {M.}~\bibnamefont {Schultze}}, \bibinfo {author} {\bibfnamefont {M.}~\bibnamefont {Fie{\ss}}}, \bibinfo {author} {\bibfnamefont {N.}~\bibnamefont {Karpowicz}}, \bibinfo {author} {\bibfnamefont {J.}~\bibnamefont {Gagnon}}, \bibinfo {author} {\bibfnamefont {M.}~\bibnamefont {Korbman}}, \bibinfo {author} {\bibfnamefont {M.}~\bibnamefont {Hofstetter}}, \bibinfo {author} {\bibfnamefont {S.}~\bibnamefont {Neppl}}, \bibinfo {author} {\bibfnamefont {A.~L.}\ \bibnamefont {Cavalieri}}, \bibinfo {author} {\bibfnamefont {Y.}~\bibnamefont {Komninos}}, \bibinfo {author} {\bibfnamefont {T.}~\bibnamefont {Mercouris}}, \emph {et~al.},\ }\bibfield  {title} {\bibinfo {title} {Delay in photoemission},\ }\href {https://www.science.org/doi/abs/10.1126/science.1189401} {\bibfield  {journal} {\bibinfo  {journal} {science}\ }\textbf {\bibinfo {volume} {328}},\ \bibinfo {pages} {1658} (\bibinfo {year} {2010})}\BibitemShut {NoStop}%
\bibitem [{\citenamefont {Cavalieri}\ \emph {et~al.}(2007)\citenamefont {Cavalieri}, \citenamefont {M{\"u}ller}, \citenamefont {Uphues}, \citenamefont {Yakovlev}, \citenamefont {Baltu{\v{s}}ka}, \citenamefont {Horvath}, \citenamefont {Schmidt}, \citenamefont {Bl{\"u}mel}, \citenamefont {Holzwarth}, \citenamefont {Hendel} \emph {et~al.}}]{cavalieri2007attosecond}%
  \BibitemOpen
  \bibfield  {author} {\bibinfo {author} {\bibfnamefont {A.~L.}\ \bibnamefont {Cavalieri}}, \bibinfo {author} {\bibfnamefont {N.}~\bibnamefont {M{\"u}ller}}, \bibinfo {author} {\bibfnamefont {T.}~\bibnamefont {Uphues}}, \bibinfo {author} {\bibfnamefont {V.~S.}\ \bibnamefont {Yakovlev}}, \bibinfo {author} {\bibfnamefont {A.}~\bibnamefont {Baltu{\v{s}}ka}}, \bibinfo {author} {\bibfnamefont {B.}~\bibnamefont {Horvath}}, \bibinfo {author} {\bibfnamefont {B.}~\bibnamefont {Schmidt}}, \bibinfo {author} {\bibfnamefont {L.}~\bibnamefont {Bl{\"u}mel}}, \bibinfo {author} {\bibfnamefont {R.}~\bibnamefont {Holzwarth}}, \bibinfo {author} {\bibfnamefont {S.}~\bibnamefont {Hendel}}, \emph {et~al.},\ }\bibfield  {title} {\bibinfo {title} {Attosecond spectroscopy in condensed matter},\ }\href {https://www.nature.com/articles/nature06229} {\bibfield  {journal} {\bibinfo  {journal} {Nature}\ }\textbf {\bibinfo {volume} {449}},\ \bibinfo {pages} {1029} (\bibinfo {year} {2007})}\BibitemShut {NoStop}%
\bibitem [{\citenamefont {Sansone}\ \emph {et~al.}(2012)\citenamefont {Sansone}, \citenamefont {Pfeifer}, \citenamefont {Simeonidis},\ and\ \citenamefont {Kuleff}}]{sansone2012electron}%
  \BibitemOpen
  \bibfield  {author} {\bibinfo {author} {\bibfnamefont {G.}~\bibnamefont {Sansone}}, \bibinfo {author} {\bibfnamefont {T.}~\bibnamefont {Pfeifer}}, \bibinfo {author} {\bibfnamefont {K.}~\bibnamefont {Simeonidis}},\ and\ \bibinfo {author} {\bibfnamefont {A.~I.}\ \bibnamefont {Kuleff}},\ }\bibfield  {title} {\bibinfo {title} {Electron correlation in real time},\ }\href {https://chemistry-europe.onlinelibrary.wiley.com/doi/full/10.1002/cphc.201100528} {\bibfield  {journal} {\bibinfo  {journal} {ChemPhysChem}\ }\textbf {\bibinfo {volume} {13}},\ \bibinfo {pages} {661} (\bibinfo {year} {2012})}\BibitemShut {NoStop}%
\bibitem [{\citenamefont {Gaumnitz}\ \emph {et~al.}(2017)\citenamefont {Gaumnitz}, \citenamefont {Jain}, \citenamefont {Pertot}, \citenamefont {Huppert}, \citenamefont {Jordan}, \citenamefont {Ardana-Lamas},\ and\ \citenamefont {W{\"o}rner}}]{gaumnitz2017streaking}%
  \BibitemOpen
  \bibfield  {author} {\bibinfo {author} {\bibfnamefont {T.}~\bibnamefont {Gaumnitz}}, \bibinfo {author} {\bibfnamefont {A.}~\bibnamefont {Jain}}, \bibinfo {author} {\bibfnamefont {Y.}~\bibnamefont {Pertot}}, \bibinfo {author} {\bibfnamefont {M.}~\bibnamefont {Huppert}}, \bibinfo {author} {\bibfnamefont {I.}~\bibnamefont {Jordan}}, \bibinfo {author} {\bibfnamefont {F.}~\bibnamefont {Ardana-Lamas}},\ and\ \bibinfo {author} {\bibfnamefont {H.~J.}\ \bibnamefont {W{\"o}rner}},\ }\bibfield  {title} {\bibinfo {title} {Streaking of 43-attosecond soft-x-ray pulses generated by a passively cep-stable mid-infrared driver},\ }\href {https://opg.optica.org/oe/fulltext.cfm?uri=oe-25-22-27506&id=375881} {\bibfield  {journal} {\bibinfo  {journal} {Optics express}\ }\textbf {\bibinfo {volume} {25}},\ \bibinfo {pages} {27506} (\bibinfo {year} {2017})}\BibitemShut {NoStop}%
\bibitem [{\citenamefont {Corkum}(1993)}]{corkum1993plasma}%
  \BibitemOpen
  \bibfield  {author} {\bibinfo {author} {\bibfnamefont {P.~B.}\ \bibnamefont {Corkum}},\ }\bibfield  {title} {\bibinfo {title} {Plasma perspective on strong field multiphoton ionization},\ }\href {https://journals.aps.org/prl/abstract/10.1103/PhysRevLett.71.1994} {\bibfield  {journal} {\bibinfo  {journal} {Physical review letters}\ }\textbf {\bibinfo {volume} {71}},\ \bibinfo {pages} {1994} (\bibinfo {year} {1993})}\BibitemShut {NoStop}%
\bibitem [{\citenamefont {Schafer}\ \emph {et~al.}(1993)\citenamefont {Schafer}, \citenamefont {Yang}, \citenamefont {DiMauro},\ and\ \citenamefont {Kulander}}]{schafer1993above}%
  \BibitemOpen
  \bibfield  {author} {\bibinfo {author} {\bibfnamefont {K.}~\bibnamefont {Schafer}}, \bibinfo {author} {\bibfnamefont {B.}~\bibnamefont {Yang}}, \bibinfo {author} {\bibfnamefont {L.}~\bibnamefont {DiMauro}},\ and\ \bibinfo {author} {\bibfnamefont {K.}~\bibnamefont {Kulander}},\ }\bibfield  {title} {\bibinfo {title} {Above threshold ionization beyond the high harmonic cutoff},\ }\href {https://journals.aps.org/prl/abstract/10.1103/PhysRevLett.70.1599} {\bibfield  {journal} {\bibinfo  {journal} {Physical review letters}\ }\textbf {\bibinfo {volume} {70}},\ \bibinfo {pages} {1599} (\bibinfo {year} {1993})}\BibitemShut {NoStop}%
\bibitem [{\citenamefont {Lewenstein}\ \emph {et~al.}(1994)\citenamefont {Lewenstein}, \citenamefont {Balcou}, \citenamefont {Ivanov}, \citenamefont {L’huillier},\ and\ \citenamefont {Corkum}}]{lewenstein1994theory}%
  \BibitemOpen
  \bibfield  {author} {\bibinfo {author} {\bibfnamefont {M.}~\bibnamefont {Lewenstein}}, \bibinfo {author} {\bibfnamefont {P.}~\bibnamefont {Balcou}}, \bibinfo {author} {\bibfnamefont {M.~Y.}\ \bibnamefont {Ivanov}}, \bibinfo {author} {\bibfnamefont {A.}~\bibnamefont {L’huillier}},\ and\ \bibinfo {author} {\bibfnamefont {P.~B.}\ \bibnamefont {Corkum}},\ }\bibfield  {title} {\bibinfo {title} {Theory of high-harmonic generation by low-frequency laser fields},\ }\href {https://journals.aps.org/pra/abstract/10.1103/PhysRevA.49.2117} {\bibfield  {journal} {\bibinfo  {journal} {Physical Review A}\ }\textbf {\bibinfo {volume} {49}},\ \bibinfo {pages} {2117} (\bibinfo {year} {1994})}\BibitemShut {NoStop}%
\bibitem [{\citenamefont {Ghimire}\ \emph {et~al.}(2011)\citenamefont {Ghimire}, \citenamefont {DiChiara}, \citenamefont {Sistrunk}, \citenamefont {Agostini}, \citenamefont {DiMauro},\ and\ \citenamefont {Reis}}]{ghimire2011observation}%
  \BibitemOpen
  \bibfield  {author} {\bibinfo {author} {\bibfnamefont {S.}~\bibnamefont {Ghimire}}, \bibinfo {author} {\bibfnamefont {A.~D.}\ \bibnamefont {DiChiara}}, \bibinfo {author} {\bibfnamefont {E.}~\bibnamefont {Sistrunk}}, \bibinfo {author} {\bibfnamefont {P.}~\bibnamefont {Agostini}}, \bibinfo {author} {\bibfnamefont {L.~F.}\ \bibnamefont {DiMauro}},\ and\ \bibinfo {author} {\bibfnamefont {D.~A.}\ \bibnamefont {Reis}},\ }\bibfield  {title} {\bibinfo {title} {Observation of high-order harmonic generation in a bulk crystal},\ }\href {https://www.nature.com/articles/nphys1847} {\bibfield  {journal} {\bibinfo  {journal} {Nature physics}\ }\textbf {\bibinfo {volume} {7}},\ \bibinfo {pages} {138} (\bibinfo {year} {2011})}\BibitemShut {NoStop}%
\bibitem [{\citenamefont {Ghimire}\ \emph {et~al.}(2012)\citenamefont {Ghimire}, \citenamefont {DiChiara}, \citenamefont {Sistrunk}, \citenamefont {Ndabashimiye}, \citenamefont {Szafruga}, \citenamefont {Mohammad}, \citenamefont {Agostini}, \citenamefont {DiMauro},\ and\ \citenamefont {Reis}}]{ghimire2012generation}%
  \BibitemOpen
  \bibfield  {author} {\bibinfo {author} {\bibfnamefont {S.}~\bibnamefont {Ghimire}}, \bibinfo {author} {\bibfnamefont {A.~D.}\ \bibnamefont {DiChiara}}, \bibinfo {author} {\bibfnamefont {E.}~\bibnamefont {Sistrunk}}, \bibinfo {author} {\bibfnamefont {G.}~\bibnamefont {Ndabashimiye}}, \bibinfo {author} {\bibfnamefont {U.~B.}\ \bibnamefont {Szafruga}}, \bibinfo {author} {\bibfnamefont {A.}~\bibnamefont {Mohammad}}, \bibinfo {author} {\bibfnamefont {P.}~\bibnamefont {Agostini}}, \bibinfo {author} {\bibfnamefont {L.~F.}\ \bibnamefont {DiMauro}},\ and\ \bibinfo {author} {\bibfnamefont {D.~A.}\ \bibnamefont {Reis}},\ }\bibfield  {title} {\bibinfo {title} {Generation and propagation of high-order harmonics in crystals},\ }\href {https://journals.aps.org/pra/abstract/10.1103/PhysRevA.85.043836} {\bibfield  {journal} {\bibinfo  {journal} {Physical Review A}\ }\textbf {\bibinfo {volume} {85}},\ \bibinfo {pages} {043836} (\bibinfo {year} {2012})}\BibitemShut {NoStop}%
\bibitem [{\citenamefont {Ndabashimiye}\ \emph {et~al.}(2016)\citenamefont {Ndabashimiye}, \citenamefont {Ghimire}, \citenamefont {Wu}, \citenamefont {Browne}, \citenamefont {Schafer}, \citenamefont {Gaarde},\ and\ \citenamefont {Reis}}]{ndabashimiye2016solid}%
  \BibitemOpen
  \bibfield  {author} {\bibinfo {author} {\bibfnamefont {G.}~\bibnamefont {Ndabashimiye}}, \bibinfo {author} {\bibfnamefont {S.}~\bibnamefont {Ghimire}}, \bibinfo {author} {\bibfnamefont {M.}~\bibnamefont {Wu}}, \bibinfo {author} {\bibfnamefont {D.~A.}\ \bibnamefont {Browne}}, \bibinfo {author} {\bibfnamefont {K.~J.}\ \bibnamefont {Schafer}}, \bibinfo {author} {\bibfnamefont {M.~B.}\ \bibnamefont {Gaarde}},\ and\ \bibinfo {author} {\bibfnamefont {D.~A.}\ \bibnamefont {Reis}},\ }\bibfield  {title} {\bibinfo {title} {Solid-state harmonics beyond the atomic limit},\ }\href {https://www.nature.com/articles/nature17660} {\bibfield  {journal} {\bibinfo  {journal} {Nature}\ }\textbf {\bibinfo {volume} {534}},\ \bibinfo {pages} {520} (\bibinfo {year} {2016})}\BibitemShut {NoStop}%
\bibitem [{\citenamefont {Liu}\ \emph {et~al.}(2017)\citenamefont {Liu}, \citenamefont {Li}, \citenamefont {You}, \citenamefont {Ghimire}, \citenamefont {Heinz},\ and\ \citenamefont {Reis}}]{liu2017high}%
  \BibitemOpen
  \bibfield  {author} {\bibinfo {author} {\bibfnamefont {H.}~\bibnamefont {Liu}}, \bibinfo {author} {\bibfnamefont {Y.}~\bibnamefont {Li}}, \bibinfo {author} {\bibfnamefont {Y.~S.}\ \bibnamefont {You}}, \bibinfo {author} {\bibfnamefont {S.}~\bibnamefont {Ghimire}}, \bibinfo {author} {\bibfnamefont {T.~F.}\ \bibnamefont {Heinz}},\ and\ \bibinfo {author} {\bibfnamefont {D.~A.}\ \bibnamefont {Reis}},\ }\bibfield  {title} {\bibinfo {title} {High-harmonic generation from an atomically thin semiconductor},\ }\href {https://www.nature.com/articles/nphys3946} {\bibfield  {journal} {\bibinfo  {journal} {Nature Physics}\ }\textbf {\bibinfo {volume} {13}},\ \bibinfo {pages} {262} (\bibinfo {year} {2017})}\BibitemShut {NoStop}%
\bibitem [{\citenamefont {You}\ \emph {et~al.}(2017)\citenamefont {You}, \citenamefont {Wu}, \citenamefont {Yin}, \citenamefont {Chew}, \citenamefont {Ren}, \citenamefont {Gholam-Mirzaei}, \citenamefont {Browne}, \citenamefont {Chini}, \citenamefont {Chang}, \citenamefont {Schafer} \emph {et~al.}}]{you2017laser}%
  \BibitemOpen
  \bibfield  {author} {\bibinfo {author} {\bibfnamefont {Y.~S.}\ \bibnamefont {You}}, \bibinfo {author} {\bibfnamefont {M.}~\bibnamefont {Wu}}, \bibinfo {author} {\bibfnamefont {Y.}~\bibnamefont {Yin}}, \bibinfo {author} {\bibfnamefont {A.}~\bibnamefont {Chew}}, \bibinfo {author} {\bibfnamefont {X.}~\bibnamefont {Ren}}, \bibinfo {author} {\bibfnamefont {S.}~\bibnamefont {Gholam-Mirzaei}}, \bibinfo {author} {\bibfnamefont {D.~A.}\ \bibnamefont {Browne}}, \bibinfo {author} {\bibfnamefont {M.}~\bibnamefont {Chini}}, \bibinfo {author} {\bibfnamefont {Z.}~\bibnamefont {Chang}}, \bibinfo {author} {\bibfnamefont {K.~J.}\ \bibnamefont {Schafer}}, \emph {et~al.},\ }\bibfield  {title} {\bibinfo {title} {Laser waveform control of extreme ultraviolet high harmonics from solids},\ }\href {https://opg.optica.org/ol/abstract.cfm?uri=ol-42-9-1816} {\bibfield  {journal} {\bibinfo  {journal} {Optics letters}\ }\textbf {\bibinfo {volume} {42}},\ \bibinfo {pages} {1816} (\bibinfo {year} {2017})}\BibitemShut {NoStop}%
\bibitem [{\citenamefont {Wu}\ \emph {et~al.}(2016)\citenamefont {Wu}, \citenamefont {Browne}, \citenamefont {Schafer},\ and\ \citenamefont {Gaarde}}]{wu2016multilevel}%
  \BibitemOpen
  \bibfield  {author} {\bibinfo {author} {\bibfnamefont {M.}~\bibnamefont {Wu}}, \bibinfo {author} {\bibfnamefont {D.~A.}\ \bibnamefont {Browne}}, \bibinfo {author} {\bibfnamefont {K.~J.}\ \bibnamefont {Schafer}},\ and\ \bibinfo {author} {\bibfnamefont {M.~B.}\ \bibnamefont {Gaarde}},\ }\bibfield  {title} {\bibinfo {title} {Multilevel perspective on high-order harmonic generation in solids},\ }\href {https://journals.aps.org/pra/abstract/10.1103/PhysRevA.94.063403} {\bibfield  {journal} {\bibinfo  {journal} {Physical Review A}\ }\textbf {\bibinfo {volume} {94}},\ \bibinfo {pages} {063403} (\bibinfo {year} {2016})}\BibitemShut {NoStop}%
\bibitem [{\citenamefont {Wu}\ \emph {et~al.}(2015)\citenamefont {Wu}, \citenamefont {Ghimire}, \citenamefont {Reis}, \citenamefont {Schafer},\ and\ \citenamefont {Gaarde}}]{wu2015high}%
  \BibitemOpen
  \bibfield  {author} {\bibinfo {author} {\bibfnamefont {M.}~\bibnamefont {Wu}}, \bibinfo {author} {\bibfnamefont {S.}~\bibnamefont {Ghimire}}, \bibinfo {author} {\bibfnamefont {D.~A.}\ \bibnamefont {Reis}}, \bibinfo {author} {\bibfnamefont {K.~J.}\ \bibnamefont {Schafer}},\ and\ \bibinfo {author} {\bibfnamefont {M.~B.}\ \bibnamefont {Gaarde}},\ }\bibfield  {title} {\bibinfo {title} {High-harmonic generation from bloch electrons in solids},\ }\href {https://journals.aps.org/pra/abstract/10.1103/PhysRevA.91.043839} {\bibfield  {journal} {\bibinfo  {journal} {Physical Review A}\ }\textbf {\bibinfo {volume} {91}},\ \bibinfo {pages} {043839} (\bibinfo {year} {2015})}\BibitemShut {NoStop}%
\bibitem [{\citenamefont {Guan}\ \emph {et~al.}(2016)\citenamefont {Guan}, \citenamefont {Zhou},\ and\ \citenamefont {Bian}}]{guan2016high}%
  \BibitemOpen
  \bibfield  {author} {\bibinfo {author} {\bibfnamefont {Z.}~\bibnamefont {Guan}}, \bibinfo {author} {\bibfnamefont {X.-X.}\ \bibnamefont {Zhou}},\ and\ \bibinfo {author} {\bibfnamefont {X.-B.}\ \bibnamefont {Bian}},\ }\bibfield  {title} {\bibinfo {title} {High-order-harmonic generation from periodic potentials driven by few-cycle laser pulses},\ }\href {https://journals.aps.org/pra/abstract/10.1103/PhysRevA.93.033852} {\bibfield  {journal} {\bibinfo  {journal} {Physical Review A}\ }\textbf {\bibinfo {volume} {93}},\ \bibinfo {pages} {033852} (\bibinfo {year} {2016})}\BibitemShut {NoStop}%
\bibitem [{\citenamefont {Ikemachi}\ \emph {et~al.}(2017)\citenamefont {Ikemachi}, \citenamefont {Shinohara}, \citenamefont {Sato}, \citenamefont {Yumoto}, \citenamefont {Kuwata-Gonokami},\ and\ \citenamefont {Ishikawa}}]{ikemachi2017trajectory}%
  \BibitemOpen
  \bibfield  {author} {\bibinfo {author} {\bibfnamefont {T.}~\bibnamefont {Ikemachi}}, \bibinfo {author} {\bibfnamefont {Y.}~\bibnamefont {Shinohara}}, \bibinfo {author} {\bibfnamefont {T.}~\bibnamefont {Sato}}, \bibinfo {author} {\bibfnamefont {J.}~\bibnamefont {Yumoto}}, \bibinfo {author} {\bibfnamefont {M.}~\bibnamefont {Kuwata-Gonokami}},\ and\ \bibinfo {author} {\bibfnamefont {K.~L.}\ \bibnamefont {Ishikawa}},\ }\bibfield  {title} {\bibinfo {title} {Trajectory analysis of high-order-harmonic generation from periodic crystals},\ }\href {https://journals.aps.org/pra/abstract/10.1103/PhysRevA.95.043416} {\bibfield  {journal} {\bibinfo  {journal} {Physical Review A}\ }\textbf {\bibinfo {volume} {95}},\ \bibinfo {pages} {043416} (\bibinfo {year} {2017})}\BibitemShut {NoStop}%
\bibitem [{\citenamefont {Tancogne-Dejean}\ \emph {et~al.}(2017{\natexlab{a}})\citenamefont {Tancogne-Dejean}, \citenamefont {M{\"u}cke}, \citenamefont {K{\"a}rtner},\ and\ \citenamefont {Rubio}}]{tancogne2017impact}%
  \BibitemOpen
  \bibfield  {author} {\bibinfo {author} {\bibfnamefont {N.}~\bibnamefont {Tancogne-Dejean}}, \bibinfo {author} {\bibfnamefont {O.~D.}\ \bibnamefont {M{\"u}cke}}, \bibinfo {author} {\bibfnamefont {F.~X.}\ \bibnamefont {K{\"a}rtner}},\ and\ \bibinfo {author} {\bibfnamefont {A.}~\bibnamefont {Rubio}},\ }\bibfield  {title} {\bibinfo {title} {Impact of the electronic band structure in high-harmonic generation spectra of solids},\ }\href {https://journals.aps.org/prl/abstract/10.1103/PhysRevLett.118.087403} {\bibfield  {journal} {\bibinfo  {journal} {Physical review letters}\ }\textbf {\bibinfo {volume} {118}},\ \bibinfo {pages} {087403} (\bibinfo {year} {2017}{\natexlab{a}})}\BibitemShut {NoStop}%
\bibitem [{\citenamefont {Tancogne-Dejean}\ \emph {et~al.}(2017{\natexlab{b}})\citenamefont {Tancogne-Dejean}, \citenamefont {M{\"u}cke}, \citenamefont {K{\"a}rtner},\ and\ \citenamefont {Rubio}}]{tancogne2017ellipticity}%
  \BibitemOpen
  \bibfield  {author} {\bibinfo {author} {\bibfnamefont {N.}~\bibnamefont {Tancogne-Dejean}}, \bibinfo {author} {\bibfnamefont {O.~D.}\ \bibnamefont {M{\"u}cke}}, \bibinfo {author} {\bibfnamefont {F.~X.}\ \bibnamefont {K{\"a}rtner}},\ and\ \bibinfo {author} {\bibfnamefont {A.}~\bibnamefont {Rubio}},\ }\bibfield  {title} {\bibinfo {title} {Ellipticity dependence of high-harmonic generation in solids originating from coupled intraband and interband dynamics},\ }\href {https://www.nature.com/articles/s41467-017-00764-5} {\bibfield  {journal} {\bibinfo  {journal} {Nature communications}\ }\textbf {\bibinfo {volume} {8}},\ \bibinfo {pages} {745} (\bibinfo {year} {2017}{\natexlab{b}})}\BibitemShut {NoStop}%
\bibitem [{\citenamefont {Tancogne-Dejean}\ and\ \citenamefont {Rubio}(2018)}]{tancogne2018atomic}%
  \BibitemOpen
  \bibfield  {author} {\bibinfo {author} {\bibfnamefont {N.}~\bibnamefont {Tancogne-Dejean}}\ and\ \bibinfo {author} {\bibfnamefont {A.}~\bibnamefont {Rubio}},\ }\bibfield  {title} {\bibinfo {title} {Atomic-like high-harmonic generation from two-dimensional materials},\ }\href {https://www.science.org/doi/10.1126/sciadv.aao5207} {\bibfield  {journal} {\bibinfo  {journal} {Science advances}\ }\textbf {\bibinfo {volume} {4}},\ \bibinfo {pages} {eaao5207} (\bibinfo {year} {2018})}\BibitemShut {NoStop}%
\bibitem [{\citenamefont {Guan}\ \emph {et~al.}(2019)\citenamefont {Guan}, \citenamefont {Lian}, \citenamefont {Hu}, \citenamefont {Liu}, \citenamefont {Zhang}, \citenamefont {Zhang},\ and\ \citenamefont {Meng}}]{guan2019cooperative}%
  \BibitemOpen
  \bibfield  {author} {\bibinfo {author} {\bibfnamefont {M.-X.}\ \bibnamefont {Guan}}, \bibinfo {author} {\bibfnamefont {C.}~\bibnamefont {Lian}}, \bibinfo {author} {\bibfnamefont {S.-Q.}\ \bibnamefont {Hu}}, \bibinfo {author} {\bibfnamefont {H.}~\bibnamefont {Liu}}, \bibinfo {author} {\bibfnamefont {S.-J.}\ \bibnamefont {Zhang}}, \bibinfo {author} {\bibfnamefont {J.}~\bibnamefont {Zhang}},\ and\ \bibinfo {author} {\bibfnamefont {S.}~\bibnamefont {Meng}},\ }\bibfield  {title} {\bibinfo {title} {Cooperative evolution of intraband and interband excitations for high-harmonic generation in strained mos 2},\ }\href {https://journals.aps.org/prb/abstract/10.1103/PhysRevB.99.184306} {\bibfield  {journal} {\bibinfo  {journal} {Physical Review B}\ }\textbf {\bibinfo {volume} {99}},\ \bibinfo {pages} {184306} (\bibinfo {year} {2019})}\BibitemShut {NoStop}%
\bibitem [{\citenamefont {Du}\ and\ \citenamefont {Bian}(2017)}]{du2017quasi}%
  \BibitemOpen
  \bibfield  {author} {\bibinfo {author} {\bibfnamefont {T.-Y.}\ \bibnamefont {Du}}\ and\ \bibinfo {author} {\bibfnamefont {X.-B.}\ \bibnamefont {Bian}},\ }\bibfield  {title} {\bibinfo {title} {Quasi-classical analysis of the dynamics of the high-order harmonic generation from solids},\ }\href {https://opg.optica.org/oe/fulltext.cfm?uri=oe-25-1-151&id=357035} {\bibfield  {journal} {\bibinfo  {journal} {Optics express}\ }\textbf {\bibinfo {volume} {25}},\ \bibinfo {pages} {151} (\bibinfo {year} {2017})}\BibitemShut {NoStop}%
\bibitem [{\citenamefont {Ghimire}\ and\ \citenamefont {Reis}(2019)}]{ghimire2019high}%
  \BibitemOpen
  \bibfield  {author} {\bibinfo {author} {\bibfnamefont {S.}~\bibnamefont {Ghimire}}\ and\ \bibinfo {author} {\bibfnamefont {D.~A.}\ \bibnamefont {Reis}},\ }\bibfield  {title} {\bibinfo {title} {High-harmonic generation from solids},\ }\href {https://www.nature.com/articles/s41567-018-0315-5} {\bibfield  {journal} {\bibinfo  {journal} {Nature physics}\ }\textbf {\bibinfo {volume} {15}},\ \bibinfo {pages} {10} (\bibinfo {year} {2019})}\BibitemShut {NoStop}%
\bibitem [{\citenamefont {Vampa}\ \emph {et~al.}(2015)\citenamefont {Vampa}, \citenamefont {Hammond}, \citenamefont {Thir{\'e}}, \citenamefont {Schmidt}, \citenamefont {L{\'e}gar{\'e}}, \citenamefont {McDonald}, \citenamefont {Brabec}, \citenamefont {Klug},\ and\ \citenamefont {Corkum}}]{vampa2015all}%
  \BibitemOpen
  \bibfield  {author} {\bibinfo {author} {\bibfnamefont {G.}~\bibnamefont {Vampa}}, \bibinfo {author} {\bibfnamefont {T.}~\bibnamefont {Hammond}}, \bibinfo {author} {\bibfnamefont {N.}~\bibnamefont {Thir{\'e}}}, \bibinfo {author} {\bibfnamefont {B.}~\bibnamefont {Schmidt}}, \bibinfo {author} {\bibfnamefont {F.}~\bibnamefont {L{\'e}gar{\'e}}}, \bibinfo {author} {\bibfnamefont {C.}~\bibnamefont {McDonald}}, \bibinfo {author} {\bibfnamefont {T.}~\bibnamefont {Brabec}}, \bibinfo {author} {\bibfnamefont {D.}~\bibnamefont {Klug}},\ and\ \bibinfo {author} {\bibfnamefont {P.}~\bibnamefont {Corkum}},\ }\bibfield  {title} {\bibinfo {title} {All-optical reconstruction of crystal band structure},\ }\href {https://journals.aps.org/prl/abstract/10.1103/PhysRevLett.115.193603} {\bibfield  {journal} {\bibinfo  {journal} {Physical review letters}\ }\textbf {\bibinfo {volume} {115}},\ \bibinfo {pages} {193603} (\bibinfo {year} {2015})}\BibitemShut {NoStop}%
\bibitem [{\citenamefont {Hu}\ \emph {et~al.}(2024{\natexlab{a}})\citenamefont {Hu}, \citenamefont {Chen}, \citenamefont {Du},\ and\ \citenamefont {Meng}}]{hu2024solid}%
  \BibitemOpen
  \bibfield  {author} {\bibinfo {author} {\bibfnamefont {S.-Q.}\ \bibnamefont {Hu}}, \bibinfo {author} {\bibfnamefont {D.-Q.}\ \bibnamefont {Chen}}, \bibinfo {author} {\bibfnamefont {L.-L.}\ \bibnamefont {Du}},\ and\ \bibinfo {author} {\bibfnamefont {S.}~\bibnamefont {Meng}},\ }\bibfield  {title} {\bibinfo {title} {Solid-state high harmonic spectroscopy for all-optical band structure probing of high-pressure quantum states},\ }\href {https://www.pnas.org/doi/abs/10.1073/pnas.2316775121} {\bibfield  {journal} {\bibinfo  {journal} {Proceedings of the National Academy of Sciences}\ }\textbf {\bibinfo {volume} {121}},\ \bibinfo {pages} {e2316775121} (\bibinfo {year} {2024}{\natexlab{a}})}\BibitemShut {NoStop}%
\bibitem [{\citenamefont {Luu}\ and\ \citenamefont {W{\"o}rner}(2018)}]{luu2018measurement}%
  \BibitemOpen
  \bibfield  {author} {\bibinfo {author} {\bibfnamefont {T.~T.}\ \bibnamefont {Luu}}\ and\ \bibinfo {author} {\bibfnamefont {H.~J.}\ \bibnamefont {W{\"o}rner}},\ }\bibfield  {title} {\bibinfo {title} {Measurement of the berry curvature of solids using high-harmonic spectroscopy},\ }\href {https://www.nature.com/articles/s41467-018-03397-4} {\bibfield  {journal} {\bibinfo  {journal} {Nature communications}\ }\textbf {\bibinfo {volume} {9}},\ \bibinfo {pages} {916} (\bibinfo {year} {2018})}\BibitemShut {NoStop}%
\bibitem [{\citenamefont {Lakhotia}\ \emph {et~al.}(2020)\citenamefont {Lakhotia}, \citenamefont {Kim}, \citenamefont {Zhan}, \citenamefont {Hu}, \citenamefont {Meng},\ and\ \citenamefont {Goulielmakis}}]{lakhotia2020laser}%
  \BibitemOpen
  \bibfield  {author} {\bibinfo {author} {\bibfnamefont {H.}~\bibnamefont {Lakhotia}}, \bibinfo {author} {\bibfnamefont {H.}~\bibnamefont {Kim}}, \bibinfo {author} {\bibfnamefont {M.}~\bibnamefont {Zhan}}, \bibinfo {author} {\bibfnamefont {S.}~\bibnamefont {Hu}}, \bibinfo {author} {\bibfnamefont {S.}~\bibnamefont {Meng}},\ and\ \bibinfo {author} {\bibfnamefont {E.}~\bibnamefont {Goulielmakis}},\ }\bibfield  {title} {\bibinfo {title} {Laser picoscopy of valence electrons in solids},\ }\href {https://www.nature.com/articles/s41586-020-2429-z} {\bibfield  {journal} {\bibinfo  {journal} {Nature}\ }\textbf {\bibinfo {volume} {583}},\ \bibinfo {pages} {55} (\bibinfo {year} {2020})}\BibitemShut {NoStop}%
\bibitem [{\citenamefont {Hu}\ \emph {et~al.}(2024{\natexlab{b}})\citenamefont {Hu}, \citenamefont {Zhao}, \citenamefont {Liu}, \citenamefont {Chen}, \citenamefont {Chen}, \citenamefont {Zhang},\ and\ \citenamefont {Meng}}]{hu2024phonon}%
  \BibitemOpen
  \bibfield  {author} {\bibinfo {author} {\bibfnamefont {S.-Q.}\ \bibnamefont {Hu}}, \bibinfo {author} {\bibfnamefont {H.}~\bibnamefont {Zhao}}, \bibinfo {author} {\bibfnamefont {X.-B.}\ \bibnamefont {Liu}}, \bibinfo {author} {\bibfnamefont {Q.}~\bibnamefont {Chen}}, \bibinfo {author} {\bibfnamefont {D.-Q.}\ \bibnamefont {Chen}}, \bibinfo {author} {\bibfnamefont {X.-Y.}\ \bibnamefont {Zhang}},\ and\ \bibinfo {author} {\bibfnamefont {S.}~\bibnamefont {Meng}},\ }\bibfield  {title} {\bibinfo {title} {Phonon-coupled high-harmonic generation for exploring nonadiabatic electron-phonon interactions},\ }\href {https://journals.aps.org/prl/abstract/10.1103/PhysRevLett.133.156901} {\bibfield  {journal} {\bibinfo  {journal} {Physical Review Letters}\ }\textbf {\bibinfo {volume} {133}},\ \bibinfo {pages} {156901} (\bibinfo {year} {2024}{\natexlab{b}})}\BibitemShut {NoStop}%
\bibitem [{\citenamefont {Li}\ \emph {et~al.}(2020)\citenamefont {Li}, \citenamefont {Lu}, \citenamefont {Chew}, \citenamefont {Han}, \citenamefont {Li}, \citenamefont {Wu}, \citenamefont {Wang}, \citenamefont {Ghimire},\ and\ \citenamefont {Chang}}]{li2020attosecond}%
  \BibitemOpen
  \bibfield  {author} {\bibinfo {author} {\bibfnamefont {J.}~\bibnamefont {Li}}, \bibinfo {author} {\bibfnamefont {J.}~\bibnamefont {Lu}}, \bibinfo {author} {\bibfnamefont {A.}~\bibnamefont {Chew}}, \bibinfo {author} {\bibfnamefont {S.}~\bibnamefont {Han}}, \bibinfo {author} {\bibfnamefont {J.}~\bibnamefont {Li}}, \bibinfo {author} {\bibfnamefont {Y.}~\bibnamefont {Wu}}, \bibinfo {author} {\bibfnamefont {H.}~\bibnamefont {Wang}}, \bibinfo {author} {\bibfnamefont {S.}~\bibnamefont {Ghimire}},\ and\ \bibinfo {author} {\bibfnamefont {Z.}~\bibnamefont {Chang}},\ }\bibfield  {title} {\bibinfo {title} {Attosecond science based on high harmonic generation from gases and solids},\ }\href {https://www.nature.com/articles/s41467-020-16480-6} {\bibfield  {journal} {\bibinfo  {journal} {Nature Communications}\ }\textbf {\bibinfo {volume} {11}},\ \bibinfo {pages} {2748} (\bibinfo {year} {2020})}\BibitemShut {NoStop}%
\bibitem [{\citenamefont {Corkum}\ and\ \citenamefont {Krausz}(2007)}]{corkum2007attosecond}%
  \BibitemOpen
  \bibfield  {author} {\bibinfo {author} {\bibfnamefont {P.~B.}\ \bibnamefont {Corkum}}\ and\ \bibinfo {author} {\bibfnamefont {F.}~\bibnamefont {Krausz}},\ }\bibfield  {title} {\bibinfo {title} {Attosecond science},\ }\href {https://www.nature.com/articles/nphys620} {\bibfield  {journal} {\bibinfo  {journal} {Nature physics}\ }\textbf {\bibinfo {volume} {3}},\ \bibinfo {pages} {381} (\bibinfo {year} {2007})}\BibitemShut {NoStop}%
\bibitem [{\citenamefont {Chini}\ \emph {et~al.}(2014)\citenamefont {Chini}, \citenamefont {Zhao},\ and\ \citenamefont {Chang}}]{chini2014generation}%
  \BibitemOpen
  \bibfield  {author} {\bibinfo {author} {\bibfnamefont {M.}~\bibnamefont {Chini}}, \bibinfo {author} {\bibfnamefont {K.}~\bibnamefont {Zhao}},\ and\ \bibinfo {author} {\bibfnamefont {Z.}~\bibnamefont {Chang}},\ }\bibfield  {title} {\bibinfo {title} {The generation, characterization and applications of broadband isolated attosecond pulses},\ }\href {https://www.nature.com/articles/nphoton.2013.362} {\bibfield  {journal} {\bibinfo  {journal} {Nature Photonics}\ }\textbf {\bibinfo {volume} {8}},\ \bibinfo {pages} {178} (\bibinfo {year} {2014})}\BibitemShut {NoStop}%
\bibitem [{\citenamefont {Krausz}\ and\ \citenamefont {Stockman}(2014)}]{krausz2014attosecond}%
  \BibitemOpen
  \bibfield  {author} {\bibinfo {author} {\bibfnamefont {F.}~\bibnamefont {Krausz}}\ and\ \bibinfo {author} {\bibfnamefont {M.~I.}\ \bibnamefont {Stockman}},\ }\bibfield  {title} {\bibinfo {title} {Attosecond metrology: from electron capture to future signal processing},\ }\href {https://www.nature.com/articles/nphoton.2014.28} {\bibfield  {journal} {\bibinfo  {journal} {Nature Photonics}\ }\textbf {\bibinfo {volume} {8}},\ \bibinfo {pages} {205} (\bibinfo {year} {2014})}\BibitemShut {NoStop}%
\bibitem [{\citenamefont {Calegari}\ \emph {et~al.}(2016)\citenamefont {Calegari}, \citenamefont {Sansone}, \citenamefont {Stagira}, \citenamefont {Vozzi},\ and\ \citenamefont {Nisoli}}]{calegari2016advances}%
  \BibitemOpen
  \bibfield  {author} {\bibinfo {author} {\bibfnamefont {F.}~\bibnamefont {Calegari}}, \bibinfo {author} {\bibfnamefont {G.}~\bibnamefont {Sansone}}, \bibinfo {author} {\bibfnamefont {S.}~\bibnamefont {Stagira}}, \bibinfo {author} {\bibfnamefont {C.}~\bibnamefont {Vozzi}},\ and\ \bibinfo {author} {\bibfnamefont {M.}~\bibnamefont {Nisoli}},\ }\bibfield  {title} {\bibinfo {title} {Advances in attosecond science},\ }\href {https://iopscience.iop.org/article/10.1088/0953-4075/49/6/062001/meta} {\bibfield  {journal} {\bibinfo  {journal} {Journal of Physics B: Atomic, Molecular and Optical Physics}\ }\textbf {\bibinfo {volume} {49}},\ \bibinfo {pages} {062001} (\bibinfo {year} {2016})}\BibitemShut {NoStop}%
\bibitem [{\citenamefont {Li}\ \emph {et~al.}(2017)\citenamefont {Li}, \citenamefont {Zhang}, \citenamefont {Yue}, \citenamefont {Wu}, \citenamefont {Hu},\ and\ \citenamefont {Du}}]{li2017enhancement}%
  \BibitemOpen
  \bibfield  {author} {\bibinfo {author} {\bibfnamefont {J.-B.}\ \bibnamefont {Li}}, \bibinfo {author} {\bibfnamefont {X.}~\bibnamefont {Zhang}}, \bibinfo {author} {\bibfnamefont {S.-J.}\ \bibnamefont {Yue}}, \bibinfo {author} {\bibfnamefont {H.-M.}\ \bibnamefont {Wu}}, \bibinfo {author} {\bibfnamefont {B.-T.}\ \bibnamefont {Hu}},\ and\ \bibinfo {author} {\bibfnamefont {H.-C.}\ \bibnamefont {Du}},\ }\bibfield  {title} {\bibinfo {title} {Enhancement of the second plateau in solid high-order harmonic spectra by the two-color fields},\ }\href {https://opg.optica.org/oe/fulltext.cfm?uri=oe-25-16-18603&id=370121} {\bibfield  {journal} {\bibinfo  {journal} {Optics Express}\ }\textbf {\bibinfo {volume} {25}},\ \bibinfo {pages} {18603} (\bibinfo {year} {2017})}\BibitemShut {NoStop}%
\bibitem [{\citenamefont {Kruchinin}\ \emph {et~al.}(2018)\citenamefont {Kruchinin}, \citenamefont {Krausz},\ and\ \citenamefont {Yakovlev}}]{kruchinin2018colloquium}%
  \BibitemOpen
  \bibfield  {author} {\bibinfo {author} {\bibfnamefont {S.~Y.}\ \bibnamefont {Kruchinin}}, \bibinfo {author} {\bibfnamefont {F.}~\bibnamefont {Krausz}},\ and\ \bibinfo {author} {\bibfnamefont {V.~S.}\ \bibnamefont {Yakovlev}},\ }\bibfield  {title} {\bibinfo {title} {Colloquium: Strong-field phenomena in periodic systems},\ }\href {https://journals.aps.org/rmp/abstract/10.1103/RevModPhys.90.021002} {\bibfield  {journal} {\bibinfo  {journal} {Reviews of Modern Physics}\ }\textbf {\bibinfo {volume} {90}},\ \bibinfo {pages} {021002} (\bibinfo {year} {2018})}\BibitemShut {NoStop}%
\bibitem [{\citenamefont {Runge}\ and\ \citenamefont {Gross}(1984)}]{runge1984density}%
  \BibitemOpen
  \bibfield  {author} {\bibinfo {author} {\bibfnamefont {E.}~\bibnamefont {Runge}}\ and\ \bibinfo {author} {\bibfnamefont {E.~K.}\ \bibnamefont {Gross}},\ }\bibfield  {title} {\bibinfo {title} {Density-functional theory for time-dependent systems},\ }\href {https://journals.aps.org/prl/abstract/10.1103/PhysRevLett.52.997} {\bibfield  {journal} {\bibinfo  {journal} {Physical review letters}\ }\textbf {\bibinfo {volume} {52}},\ \bibinfo {pages} {997} (\bibinfo {year} {1984})}\BibitemShut {NoStop}%
\bibitem [{\citenamefont {Van~Leeuwen}(1998)}]{van1998causality}%
  \BibitemOpen
  \bibfield  {author} {\bibinfo {author} {\bibfnamefont {R.}~\bibnamefont {Van~Leeuwen}},\ }\bibfield  {title} {\bibinfo {title} {Causality and symmetry in time-dependent density-functional theory},\ }\href {https://journals.aps.org/prl/abstract/10.1103/PhysRevLett.80.1280} {\bibfield  {journal} {\bibinfo  {journal} {Physical review letters}\ }\textbf {\bibinfo {volume} {80}},\ \bibinfo {pages} {1280} (\bibinfo {year} {1998})}\BibitemShut {NoStop}%
\bibitem [{\citenamefont {Wu}\ \emph {et~al.}(2021)\citenamefont {Wu}, \citenamefont {Liang}, \citenamefont {Kong}, \citenamefont {Gong},\ and\ \citenamefont {Peng}}]{wu2021enhancement}%
  \BibitemOpen
  \bibfield  {author} {\bibinfo {author} {\bibfnamefont {X.-Y.}\ \bibnamefont {Wu}}, \bibinfo {author} {\bibfnamefont {H.}~\bibnamefont {Liang}}, \bibinfo {author} {\bibfnamefont {X.-S.}\ \bibnamefont {Kong}}, \bibinfo {author} {\bibfnamefont {Q.}~\bibnamefont {Gong}},\ and\ \bibinfo {author} {\bibfnamefont {L.-Y.}\ \bibnamefont {Peng}},\ }\bibfield  {title} {\bibinfo {title} {Enhancement of high-order harmonic generation in two-dimensional materials by plasmonic fields},\ }\href {https://journals.aps.org/pra/abstract/10.1103/PhysRevA.103.043117} {\bibfield  {journal} {\bibinfo  {journal} {Physical Review A}\ }\textbf {\bibinfo {volume} {103}},\ \bibinfo {pages} {043117} (\bibinfo {year} {2021})}\BibitemShut {NoStop}%
\bibitem [{\citenamefont {Guan}\ \emph {et~al.}(2020)\citenamefont {Guan}, \citenamefont {Hu}, \citenamefont {Zhao}, \citenamefont {Lian},\ and\ \citenamefont {Meng}}]{guan2020toward}%
  \BibitemOpen
  \bibfield  {author} {\bibinfo {author} {\bibfnamefont {M.}~\bibnamefont {Guan}}, \bibinfo {author} {\bibfnamefont {S.}~\bibnamefont {Hu}}, \bibinfo {author} {\bibfnamefont {H.}~\bibnamefont {Zhao}}, \bibinfo {author} {\bibfnamefont {C.}~\bibnamefont {Lian}},\ and\ \bibinfo {author} {\bibfnamefont {S.}~\bibnamefont {Meng}},\ }\bibfield  {title} {\bibinfo {title} {Toward attosecond control of electron dynamics in two-dimensional materials},\ }\href {https://pubs.aip.org/aip/apl/article/116/4/043101/1078174/Toward-attosecond-control-of-electron-dynamics-in} {\bibfield  {journal} {\bibinfo  {journal} {Applied Physics Letters}\ }\textbf {\bibinfo {volume} {116}} (\bibinfo {year} {2020})}\BibitemShut {NoStop}%
\bibitem [{\citenamefont {Nourbakhsh}\ \emph {et~al.}(2021)\citenamefont {Nourbakhsh}, \citenamefont {Tancogne-Dejean}, \citenamefont {Merdji},\ and\ \citenamefont {Rubio}}]{nourbakhsh2021high}%
  \BibitemOpen
  \bibfield  {author} {\bibinfo {author} {\bibfnamefont {Z.}~\bibnamefont {Nourbakhsh}}, \bibinfo {author} {\bibfnamefont {N.}~\bibnamefont {Tancogne-Dejean}}, \bibinfo {author} {\bibfnamefont {H.}~\bibnamefont {Merdji}},\ and\ \bibinfo {author} {\bibfnamefont {A.}~\bibnamefont {Rubio}},\ }\bibfield  {title} {\bibinfo {title} {High harmonics and isolated attosecond pulses from mgo},\ }\href {https://journals.aps.org/prapplied/abstract/10.1103/PhysRevApplied.15.014013} {\bibfield  {journal} {\bibinfo  {journal} {Physical Review Applied}\ }\textbf {\bibinfo {volume} {15}},\ \bibinfo {pages} {014013} (\bibinfo {year} {2021})}\BibitemShut {NoStop}%
\bibitem [{\citenamefont {Sadeghifaraz}\ \emph {et~al.}(2022)\citenamefont {Sadeghifaraz}, \citenamefont {Irani},\ and\ \citenamefont {Monfared}}]{sadeghifaraz2022efficient}%
  \BibitemOpen
  \bibfield  {author} {\bibinfo {author} {\bibfnamefont {A.}~\bibnamefont {Sadeghifaraz}}, \bibinfo {author} {\bibfnamefont {E.}~\bibnamefont {Irani}},\ and\ \bibinfo {author} {\bibfnamefont {M.}~\bibnamefont {Monfared}},\ }\bibfield  {title} {\bibinfo {title} {Efficient attosecond pulse generation from ws2 semiconductor by tailoring the driving laser pulse},\ }\href {https://www.sciencedirect.com/science/article/pii/S0030401822001869} {\bibfield  {journal} {\bibinfo  {journal} {Optics Communications}\ }\textbf {\bibinfo {volume} {516}},\ \bibinfo {pages} {128226} (\bibinfo {year} {2022})}\BibitemShut {NoStop}%
\bibitem [{\citenamefont {Ullrich}(2011)}]{ullrich2011time}%
  \BibitemOpen
  \bibfield  {author} {\bibinfo {author} {\bibfnamefont {C.~A.}\ \bibnamefont {Ullrich}},\ }\href {https://books.google.com.sg/books?hl=zh-CN&lr=&id=pLycQCW-2BwC&oi=fnd&pg=PP1&dq=TDDFT+book&ots=V_D4POZtvO&sig=1DuNidLuFfo5uONVHpcIrGPZxdQ&redir_esc=y#v=onepage&q=TDDFT%20book&f=false} {\emph {\bibinfo {title} {Time-dependent density-functional theory: concepts and applications}}}\ (\bibinfo  {publisher} {OUP Oxford},\ \bibinfo {year} {2011})\BibitemShut {NoStop}%
\bibitem [{\citenamefont {Tancogne-Dejean}\ \emph {et~al.}(2020)\citenamefont {Tancogne-Dejean}, \citenamefont {Oliveira}, \citenamefont {Andrade}, \citenamefont {Appel}, \citenamefont {Borca}, \citenamefont {Le~Breton}, \citenamefont {Buchholz}, \citenamefont {Castro}, \citenamefont {Corni}, \citenamefont {Correa} \emph {et~al.}}]{tancogne2020octopus}%
  \BibitemOpen
  \bibfield  {author} {\bibinfo {author} {\bibfnamefont {N.}~\bibnamefont {Tancogne-Dejean}}, \bibinfo {author} {\bibfnamefont {M.~J.}\ \bibnamefont {Oliveira}}, \bibinfo {author} {\bibfnamefont {X.}~\bibnamefont {Andrade}}, \bibinfo {author} {\bibfnamefont {H.}~\bibnamefont {Appel}}, \bibinfo {author} {\bibfnamefont {C.~H.}\ \bibnamefont {Borca}}, \bibinfo {author} {\bibfnamefont {G.}~\bibnamefont {Le~Breton}}, \bibinfo {author} {\bibfnamefont {F.}~\bibnamefont {Buchholz}}, \bibinfo {author} {\bibfnamefont {A.}~\bibnamefont {Castro}}, \bibinfo {author} {\bibfnamefont {S.}~\bibnamefont {Corni}}, \bibinfo {author} {\bibfnamefont {A.~A.}\ \bibnamefont {Correa}}, \emph {et~al.},\ }\bibfield  {title} {\bibinfo {title} {Octopus, a computational framework for exploring light-driven phenomena and quantum dynamics in extended and finite systems},\ }\href {https://pubs.aip.org/aip/jcp/article/152/12/124119/954926} {\bibfield  {journal} {\bibinfo  {journal} {The Journal of chemical physics}\ }\textbf {\bibinfo {volume}
  {152}} (\bibinfo {year} {2020})}\BibitemShut {NoStop}%
\bibitem [{\citenamefont {Rossi}\ and\ \citenamefont {Kuhn}(2002)}]{rossi2002theory}%
  \BibitemOpen
  \bibfield  {author} {\bibinfo {author} {\bibfnamefont {F.}~\bibnamefont {Rossi}}\ and\ \bibinfo {author} {\bibfnamefont {T.}~\bibnamefont {Kuhn}},\ }\bibfield  {title} {\bibinfo {title} {Theory of ultrafast phenomena in photoexcited semiconductors},\ }\href {https://journals.aps.org/rmp/abstract/10.1103/RevModPhys.74.895} {\bibfield  {journal} {\bibinfo  {journal} {Reviews of Modern Physics}\ }\textbf {\bibinfo {volume} {74}},\ \bibinfo {pages} {895} (\bibinfo {year} {2002})}\BibitemShut {NoStop}%
\bibitem [{\citenamefont {Roberts}\ \emph {et~al.}(2011)\citenamefont {Roberts}, \citenamefont {Cormode}, \citenamefont {Reynolds}, \citenamefont {Newhouse-Illige}, \citenamefont {LeRoy},\ and\ \citenamefont {Sandhu}}]{roberts2011response}%
  \BibitemOpen
  \bibfield  {author} {\bibinfo {author} {\bibfnamefont {A.}~\bibnamefont {Roberts}}, \bibinfo {author} {\bibfnamefont {D.}~\bibnamefont {Cormode}}, \bibinfo {author} {\bibfnamefont {C.}~\bibnamefont {Reynolds}}, \bibinfo {author} {\bibfnamefont {T.}~\bibnamefont {Newhouse-Illige}}, \bibinfo {author} {\bibfnamefont {B.~J.}\ \bibnamefont {LeRoy}},\ and\ \bibinfo {author} {\bibfnamefont {A.~S.}\ \bibnamefont {Sandhu}},\ }\bibfield  {title} {\bibinfo {title} {Response of graphene to femtosecond high-intensity laser irradiation},\ }\href {https://pubs.aip.org/aip/apl/article/99/5/051912/374257} {\bibfield  {journal} {\bibinfo  {journal} {Applied Physics Letters}\ }\textbf {\bibinfo {volume} {99}} (\bibinfo {year} {2011})}\BibitemShut {NoStop}%
\bibitem [{\citenamefont {Kong}\ \emph {et~al.}(2022)\citenamefont {Kong}, \citenamefont {Liang}, \citenamefont {Wu}, \citenamefont {Geng}, \citenamefont {Yu},\ and\ \citenamefont {Peng}}]{kong2022manipulation}%
  \BibitemOpen
  \bibfield  {author} {\bibinfo {author} {\bibfnamefont {X.-S.}\ \bibnamefont {Kong}}, \bibinfo {author} {\bibfnamefont {H.}~\bibnamefont {Liang}}, \bibinfo {author} {\bibfnamefont {X.-Y.}\ \bibnamefont {Wu}}, \bibinfo {author} {\bibfnamefont {L.}~\bibnamefont {Geng}}, \bibinfo {author} {\bibfnamefont {W.-D.}\ \bibnamefont {Yu}},\ and\ \bibinfo {author} {\bibfnamefont {L.-Y.}\ \bibnamefont {Peng}},\ }\bibfield  {title} {\bibinfo {title} {Manipulation of the high-order harmonic generation in monolayer hexagonal boron nitride by two-color laser field},\ }\href {https://pubs.aip.org/aip/jcp/article/156/7/074701/2840845} {\bibfield  {journal} {\bibinfo  {journal} {The Journal of Chemical Physics}\ }\textbf {\bibinfo {volume} {156}},\ \bibinfo {pages} {074701} (\bibinfo {year} {2022})}\BibitemShut {NoStop}%
\bibitem [{\citenamefont {Rybkovskiy}\ \emph {et~al.}(2011)\citenamefont {Rybkovskiy}, \citenamefont {Arutyunyan}, \citenamefont {Orekhov}, \citenamefont {Gromchenko}, \citenamefont {Vorobiev}, \citenamefont {Osadchy}, \citenamefont {Salaev}, \citenamefont {Baykara}, \citenamefont {Allakhverdiev},\ and\ \citenamefont {Obraztsova}}]{rybkovskiy2011size}%
  \BibitemOpen
  \bibfield  {author} {\bibinfo {author} {\bibfnamefont {D.}~\bibnamefont {Rybkovskiy}}, \bibinfo {author} {\bibfnamefont {N.}~\bibnamefont {Arutyunyan}}, \bibinfo {author} {\bibfnamefont {A.}~\bibnamefont {Orekhov}}, \bibinfo {author} {\bibfnamefont {I.}~\bibnamefont {Gromchenko}}, \bibinfo {author} {\bibfnamefont {I.}~\bibnamefont {Vorobiev}}, \bibinfo {author} {\bibfnamefont {A.}~\bibnamefont {Osadchy}}, \bibinfo {author} {\bibfnamefont {E.~Y.}\ \bibnamefont {Salaev}}, \bibinfo {author} {\bibfnamefont {T.}~\bibnamefont {Baykara}}, \bibinfo {author} {\bibfnamefont {K.}~\bibnamefont {Allakhverdiev}},\ and\ \bibinfo {author} {\bibfnamefont {E.}~\bibnamefont {Obraztsova}},\ }\bibfield  {title} {\bibinfo {title} {Size-induced effects in gallium selenide electronic structure: The influence of interlayer interactions},\ }\href {https://journals.aps.org/prb/pdf/10.1103/PhysRevB.84.085314} {\bibfield  {journal} {\bibinfo  {journal} {Physical Review B}\ }\textbf {\bibinfo {volume} {84}},\ \bibinfo {pages} {085314}
  (\bibinfo {year} {2011})}\BibitemShut {NoStop}%
\bibitem [{\citenamefont {Elias}\ \emph {et~al.}(2019)\citenamefont {Elias}, \citenamefont {Valvin}, \citenamefont {Pelini}, \citenamefont {Summerfield}, \citenamefont {Mellor}, \citenamefont {Cheng}, \citenamefont {Eaves}, \citenamefont {Foxon}, \citenamefont {Beton}, \citenamefont {Novikov} \emph {et~al.}}]{elias2019direct}%
  \BibitemOpen
  \bibfield  {author} {\bibinfo {author} {\bibfnamefont {C.}~\bibnamefont {Elias}}, \bibinfo {author} {\bibfnamefont {P.}~\bibnamefont {Valvin}}, \bibinfo {author} {\bibfnamefont {T.}~\bibnamefont {Pelini}}, \bibinfo {author} {\bibfnamefont {A.}~\bibnamefont {Summerfield}}, \bibinfo {author} {\bibfnamefont {C.}~\bibnamefont {Mellor}}, \bibinfo {author} {\bibfnamefont {T.}~\bibnamefont {Cheng}}, \bibinfo {author} {\bibfnamefont {L.}~\bibnamefont {Eaves}}, \bibinfo {author} {\bibfnamefont {C.}~\bibnamefont {Foxon}}, \bibinfo {author} {\bibfnamefont {P.}~\bibnamefont {Beton}}, \bibinfo {author} {\bibfnamefont {S.}~\bibnamefont {Novikov}}, \emph {et~al.},\ }\bibfield  {title} {\bibinfo {title} {Direct band-gap crossover in epitaxial monolayer boron nitride},\ }\href {https://www.nature.com/articles/s41467-019-10610-5} {\bibfield  {journal} {\bibinfo  {journal} {Nature communications}\ }\textbf {\bibinfo {volume} {10}},\ \bibinfo {pages} {2639} (\bibinfo {year} {2019})}\BibitemShut {NoStop}%
\end{thebibliography}%
\clearpage

\end{document}